\documentclass[lettersize,journal]{IEEEtran}
\usepackage{amsfonts}
\usepackage{amssymb}
\usepackage{stfloats}
\usepackage{cite}
\usepackage{stfloats}
\usepackage{graphicx}
\usepackage{epstopdf}
\usepackage{psfrag}
\usepackage{subfig}
\usepackage{amsmath}
\usepackage{array}
\usepackage{multirow}
\usepackage{color}
\usepackage{booktabs}
\usepackage{url}
\usepackage{enumitem}
\usepackage{lipsum} 
\usepackage{cuted}
\usepackage{mathtools}
\usepackage{marvosym} 
\usepackage{url}
\usepackage{hyperref}
\usepackage{makecell}
\usepackage[linesnumbered,ruled,vlined]{algorithm2e}
\newtheorem{Thm}{Theorem}
\newtheorem{Lem}{Lemma}

\newtheorem{Def}{Definition}

\newtheorem{Prob}{Problem}
\newtheorem{subProb}{SubProblem}[Prob]

\begin{document}

\title{IREE Oriented Active RIS-Assisted Green communication System with Outdated CSI}

\author{Kai~Cao, Tao~Yu,    Jihong~Li , Xiaojing~Chen,~\IEEEmembership{Member, IEEE,} Yanzan~Sun,~\IEEEmembership{Member, IEEE,} Qingqing~Wu,~\IEEEmembership{Senior Member, IEEE,} 
Wen~Chen,~\IEEEmembership{Senior Member, IEEE,} 
Shunqing~Zhang,~\IEEEmembership{Senior Member, IEEE,}
\thanks{This work was supported by the National Key Research and Development Program of China under Grant 2022YFB2902304, Grant 2022YFB2902005, and the National Natural Science Foundation of China under Grant 62071284. }
\thanks{Kai Cao, Tao Yu, Jihong Li,  Xiaojing Chen, Yanzan Sun and Shunqing Zhang are with Shanghai Institute for Advanced Communication and Data Science, Key laboratory of Specialty Fiber Optics and Optical Access Networks, Shanghai University, Shanghai, 200444, China (e-mails: \{ck-09, yu\_tao, tomlijiong, shunqing, jodiechen, yanzansun\}@shu.edu.cn).

Qingqing Wu and Wen Chen are with the Department of Electronic Engineering, Shanghai Jiao Tong University, Shanghai
200240, China (e-mails: \{qingqingwu, wenchen\}@sjtu.edu.cn). 
}
\thanks{Corresponding Author: {\em Shunqing Zhang}.}
}



\maketitle

\begin{abstract}
The rapid evolution of communication technologies has spurred a growing demand for energy-efficient network architectures and performance metrics. Active Reconfigurable Intelligent Surfaces (RIS) are emerging as a key component in green network architectures. Compared to passive RIS, active RIS are equipped with amplifiers on each reflecting element, allowing them to simultaneously reflect and amplify signals, thereby overcoming the double multiplicative fading in the phase response, and improving both system coverage and performance. Additionally, the Integrated Relative Energy Efficiency (IREE) metric, as introduced in \cite{yu2022novel}, addresses the dynamic variations in traffic and capacity over time and space, enabling more energy-efficient wireless systems. Building on these advancements, this paper investigates the problem of maximizing IREE in active RIS-assisted green communication systems. However, acquiring perfect Channel State Information (CSI) in practical systems poses significant challenges and costs. To address this, we derive the average achievable rate based on outdated CSI and formulated the corresponding IREE maximization problem, which is solved by jointly optimizing beamforming at both the base station and RIS.
Given the non-convex nature of the problem, we propose an  Alternating Optimization Successive Approximation (AOSO)  algorithm. By applying quadratic transform and relaxation techniques, we simplify the original problem and alternately optimize the beamforming matrices at the base station and RIS. Furthermore, to handle the discrete constraints of the RIS reflection coefficients, we develop a successive approximation method. Experimental results validate our theoretical analysis of the algorithm's convergence , demonstrating the effectiveness of the proposed algorithm and highlighting the superiority of IREE in enhancing the performance of green communication networks.

\end{abstract}


\section{Introduction}
\IEEEPARstart{I}{n}  recent years, Reconfigurable Intelligent Surfaces (RIS) have emerged as a key technology for advancing green communications due to their simple deployment and low cost \cite{huang2019reconfigurable,lv2022ris}. By equipping reflective elements on the surface, RIS can reflect incoming signals in pre-determined directions, effectively controlling the wireless channel environment \cite{wu2019intelligent}. This capability allows RIS to reduce interference from unwanted signals, thereby improving the signal-to-noise ratio (SNR) at the user's receiver and enhancing overall system performance. Moreover, when obstructions such as severe attenuation in millimeter-wave communication systems occur between the base station and the user, RIS can establish an alternative link to mitigate this blockage. Additionally, compared to traditional relay networks, RIS does not require complex signal processing circuits or radio frequency (RF) chains, which significantly reduces power consumption \cite{9119122}. RIS also has minimal constraints on deployment locations and can be easily installed on walls, ceilings, and other structures in urban environments \cite{liu2021reconfigurable}.

However, passive RIS faces the challenge of multiplicative fading, leading to substantial path loss in the communication link from the base station to the user through the RIS \cite{long2021active,zhang2022active}. This issue becomes more pronounced at higher carrier frequencies. To address this, the authors in \cite{long2021active} proposed the active RIS, which amplifies the incident signal at the electromagnetic level before reflecting it. Similar to passive RIS, active RIS does not require complex or high-power RF chain components.

Extensive research has been conducted on active RIS. In \cite{zhi2022active}, the authors compared the performance of active and passive RIS under the same power budget, demonstrating that with an appropriate power budget and a moderate number of RIS elements, active RIS outperforms passive RIS. In \cite{lv2022ris}, an active RIS-assisted multi-antenna physical layer secure transmission scheme was proposed. Experimental results showed that active RIS can effectively counter the impact of multiplicative fading and is more energy-efficient than its passive counterpart. In \cite{ren2023transmission}, an active RIS-assisted Simultaneous Wireless Information and Power Transfer (SWIPT) system was studied, demonstrating that, under the same power budget, the active RIS-assisted SWIPT system achieves superior performance. Finally, in \cite{Fot2024}, the authors investigated the problem of maximizing energy efficiency for active RIS systems.

It is widely acknowledged that in practical RIS-assisted communication systems, obtaining accurate CSI is crucial for effective operation. Both the beamforming at the base station and the RIS depend heavily on CSI. However, acquiring perfect instantaneous CSI is challenging, particularly in millimeter-wave systems and those operating at higher frequencies, where the channel conditions fluctuate rapidly over time \cite{waqas2023effective}. 
Fortunately, the channel between the base station and the RIS can be considered quasi-static, as the positions of both the RIS and the base station remain fixed. To improve CSI accuracy, the authors in \cite{hu2021two} proposed a dual-timescale channel estimation scheme, where the channel between the base station and RIS is estimated on a large timescale, while the channel between the RIS and the user is estimated on a smaller timescale. Similarly, in \cite{peng2023two}, a two-stage uplink channel estimation strategy was developed for RIS-assisted multi-user millimeter-wave multi-antenna systems. However, to reduce channel estimation overhead, frequent updates of CSI between the base station and RIS are avoided. As a result, it is crucial to account for the effects of outdated CSI between the RIS and the user.

Currently, the performance metrics for RIS-assisted communication systems are largely limited to outage probability, user received SNR, traditional energy efficiency (EE) metrics, and similar measures. In \cite{guo2020weighted}, the authors investigate the maximization of weighted sum-rate in a RIS-aided multiuser downlink MISO system. Additionally, \cite{ma2023optimization} explores joint beamforming in a RIS-enhanced multi-user multiple-input single-output system, focusing on maximizing overall rate while minimizing power consumption. The study in \cite{ma2022active} examines the energy efficiency of downlink transmissions supported by active RIS, with experimental results indicating that deploying active RIS in wireless systems is a promising approach to achieving green communications. Furthermore, in \cite{wu2022energy}, energy-efficient power control is optimized as an objective in RIS-assisted Internet of Things communication systems. However, these metrics do not adequately align with the future development of green communications.  

This is mainly due to the rapid development of communication networks, which has led to an imbalanced traffic distribution in actual network deployments. The actual traffic demands of each user vary significantly. The primary advantage of RIS is its ability to actively improve the channel environment, providing users with substantial traffic gains. However, if we rely solely on traditional metrics like EE, extra power will be wasted when the actual traffic demand of users does not match the capacity provided by the RIS. Therefore, we need a new metric to effectively align user traffic demands with network capacity. We adopt the IREE \cite{yu2022novel}, which comprehensively considers traffic profiles and network capacity from an energy efficiency perspective. By utilizing the more comprehensive green metric of IREE, we can effectively leverage the capability of RIS to enhance network coverage and maximize the fulfillment of each user's traffic demands as much as possible.

In summary, we have investigated a problem of maximizing IREE in active RIS-aided millimeter-wave communication systems, addressing key challenges and limitations present in current research on active RIS. Our main contributions are as follows:

\begin{itemize}
    \item Acknowledging that perfect CSI is unattainable and that the CSI between the RIS and users is not estimated frequently to minimize channel overhead, we analyze the impact of outdated CSI and derive a closed-form expression for the average achievable rate in an active RIS-aided communication system. Building on this average achievable rate, we define the corresponding network IREE , which effectively aligns user traffic demands with  network capacity.
Based on this analysis, we propose an optimization framework to maximize IREE, which jointly optimizes the beamforming at both the base station and the RIS.

    \item To tackle the non-convexity introduced by the IREE metric and the inability to continuously adjust the reflection coefficients at the RIS, we propose an Alternating Optimization Successive Approximation (AOSO) algorithm to maximize IREE in active RIS-assisted millimeter-wave communication systems. Specifically, we simplify the original non-convex problem using quadratic transform and relaxation techniques, decoupling it for iterative solution via an alternating optimization method. Finally, to manage the discrete adjustments of active RIS transmission coefficients, we introduce a successive approximation approach.
    
    \item We conducte a theoretical analysis of the algorithm's convergence and complexity. Through experimental analysis, we verify that the algorithm achieves rapid convergence with low complexity. Additionally, the proposed algorithm effectively maximizes IREE while significantly mitigating the impact of outdated CSI.
Even with limited quantization levels for the RIS phase and amplitude, our approach closely approximates the ideal scenario of continuous reflection coefficient adjustment. Furthermore, experimental results demonstrate that the IREE metric enables RIS-assisted millimeter-wave communication systems to better align network capacity with user traffic demands compared to traditional EE metrics, thereby establishing its leading role in the future of green communications.

\end{itemize}

\emph{Notations:}
In this paper, we denote vectors using lowercase symbols and matrices using uppercase symbols. Specifically, $\mathbf{X}^{-1}$, $\mathbf{X}^{T}$, and $\mathbf{X}^{H}$ represent the inverse, transpose, and conjugate transpose of the matrix $\mathbf{X}$, respectively. Additionally, $\text{diag}(\mathbf{X})$ denotes a diagonal matrix formed from the diagonal elements of $\mathbf{X}$, while $\mathbf{I}_n$ represents the identity matrix of size $n \times n$.
We also use the notation $\mathbb{E}\{\cdot\}$, $\text{Tr}\{\cdot\}$, and $\text{Re}\{\cdot\}$ to indicate the expectation, trace, and real part of a complex number, respectively. The symbols $|\cdot|$ and $\|\cdot\|$ are used to denote the absolute value of a complex number and the norm of a vector, respectively. Furthermore, $\mathcal{CN}(0, \sigma^2)$ represents a vector which follows a circularly symmetric complex Gaussian distribution with zero mean and covariance matrix $\sigma^2$ .

\section{System Models and Problem Formulation}
\label{sect:system_model} 

In this section, we briefly present an active RIS-assisted multi-user multiple-input single-output (MISO) system and formulate the IREE maximization problem in what follows.

\subsection{System Model}

\begin{figure}[!t]
        \centering
        \includegraphics[width=1\columnwidth]{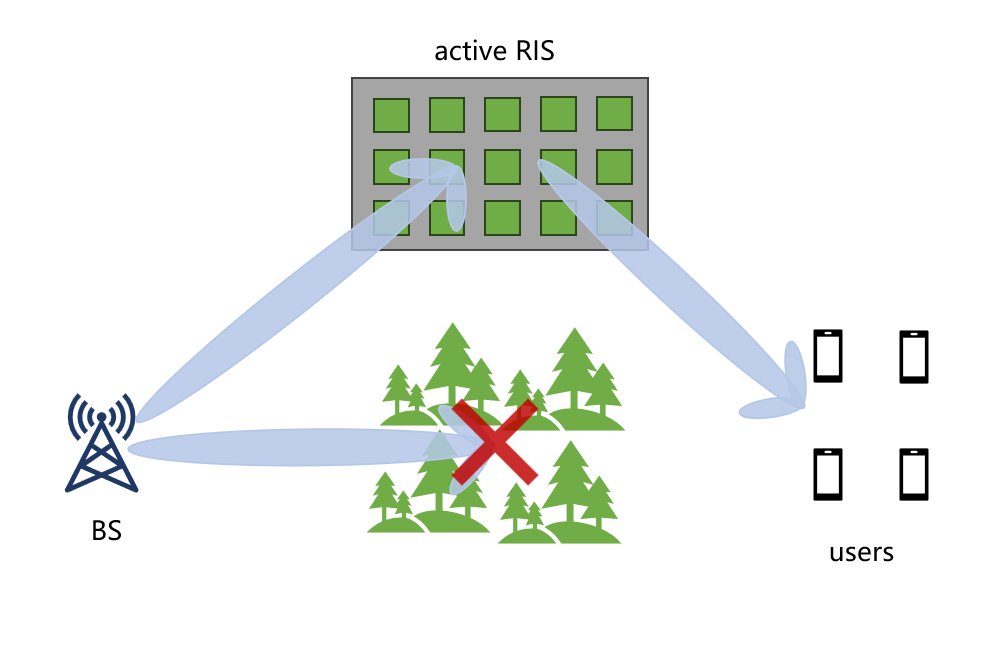}
        \caption{Schematic diagram of the communication system}
        \label{fig:RIS}
\end{figure}

We consider an active RIS-assisted downlink transmission system as shown in Fig.~\ref{fig:RIS}, where the base station with $N_t$ transmitting antennas is communicating with $K$ single-antenna users simultaneously with the help of  $N_{RIS}$-element active RIS.
Denote $\mathbf{H}_{br} \in \mathbb{C}^{N_{RIS} \times N_{t}}$ to be the channel fading coefficients for the base station(BS) to the  RIS link and $\mathbf{h}^{H}_{ru,k} \in \mathbb{C}^{1 \times N_{RIS}}$ to be the channel fading environment between the RIS and the $k$-th user, respectively.  The received signal at the $k$-th user, $y_k$ is therefore given by,
\begin{eqnarray}
\label{eq:yk}
    y_k =\mathbf{h}^{H}_{ru,k} \boldsymbol{\Theta} \mathbf{H}_{br} \sum_{k=1}^{K} \mathbf{w}_{k} {x}_{k} 
+  \mathbf{h}^{H}_{ru,k} \boldsymbol{\Theta}\mathbf{n}_{R} + n_k,
\end{eqnarray}
where $\mathbf n_{R} \sim \mathcal{CN}(0,\sigma^2_{R}\mathbf{I}_{R})$ and $n_k \sim \mathcal{CN}(0,\sigma^2_{n})$ denote the additive white Gaussian noise (AWGN) at the active RIS and the $k$-th user sides, respectively. ${x}_k  $ denotes the normalized transmit symbols for the $k$-th user. $\boldsymbol\Theta=\textrm{diag}(\boldsymbol{\theta}) \in \mathbb C^{N_{RIS} \times N_{RIS}}$ with $\boldsymbol{\theta}=[\theta_{1},\dots,\theta_{n},...,\theta_{N_{RIS}}]$ and $\mathbf{W} = [\mathbf{w}_{1}, \ldots, \mathbf{w}_{k}, \ldots, \mathbf{w}_{K}] \in \mathbb C^{N_{t} \times K}$ denote the beamforming matrices at the RIS and the BS sides, respectively. The received signal-to-interference-noise ratio (SINR) at the $k$-th user, $\gamma_k$, is equal to,
\begin{eqnarray}
\gamma_k = \frac{|\mathbf{h}_{ru,k}^H \boldsymbol\Theta \mathbf{H}_{br} \mathbf{w}_{k}|^2}{\sum^{K}_{ k' \neq k} |\mathbf{h}^H_{ru,k} \boldsymbol\Theta \mathbf{H}_{br}\mathbf{w}_{k'}|^2 + ||\mathbf{h}^H_{ru,k}\mathbf\Theta||^2\sigma^2_{R}+\sigma_n^2}.
\label{eq:snr}
\end{eqnarray}
The achievable transmission rate of the $k$-th user, $R_k\left(\mathbf \Theta,\mathbf{W}\right)$, is thus given by, $R_k\left(\mathbf \Theta,\mathbf{W}\right) = BW \times \log_2(1+\gamma_k)$ with $BW$ denoting the corresponding transmission bandwidth. 

Due to the channel outdatedness and the imperfect channel estimation schemes as specified in \cite{yang2020deep}, the estimated channel response, $\hat{\mathbf h}_{ru,k}$, and the perfect channel response, ${\mathbf h}_{ru,k}$, has the following relations,
\begin{eqnarray}
    \mathbf h_{ru,k} = \rho \hat{\mathbf h}_{ru,k}+\sqrt{1-\rho^2}\Delta{\mathbf h}_{ru,k}, \label{eq:imcsi}
\end{eqnarray}
where $\rho \in (0,1)$ denotes the correlation coefficient as defined in \cite{yang2020deep}, and $\Delta{\mathbf h}_{ru, k} \sim \mathcal{CN}(0,\sigma^2_{ru,k}\mathbf{I}_{R})$ denotes the residual estimation error correspondingly. With the above settings, the average achievable rate $\bar{R}_k$ can be obtained through the following lemma.
\begin{Lem}[Average Achievable Rate $\bar{R}_k$] \label{le:rk} The average achievable rate $\bar{R}_k$ under the imperfect channel estimation $\hat{\mathbf h}_{ru,k}$ is given by,
\begin{align}\label{eq:rk}
  \mathbb{E}\left[R_k|\hat{\mathbf h}_{ru,k}\right]    \geq \bar{R}_k \left(\mathbf{\Theta}, \mathbf{W}\right)  = BW \times  \log_2\bigg(1+ \nonumber \\
    \frac{|\rho\hat{\mathbf h}_{ru,k} \boldsymbol\Theta \mathbf{H}_{br}\mathbf{w}_{k}|^2}{\sum_{k'= 1}^K||\mathbf{Q}_{ki} \Theta \mathbf{H}_{br}\mathbf{w}_{k'}||^2+|| \mathbf{Q}_{k0}\boldsymbol\Theta||^2\sigma_{R}^2 +\sigma_n^2}\bigg),
\end{align}
where $$ \mathbf{Q}_{kk'} =
     \begin{cases} 
      \sqrt{\rho\hat{\mathbf h}_{ru,k} \hat{\mathbf h}_{ru,k}^H + (1-\rho^2)\sigma_{ru,k}^2 \mathbf{I}_{R}},  &k'\neq k, \\
      \sqrt{1-\rho^2}\sigma_{ru,k},  &k' = k.
     \end{cases} $$
\IEEEproof Please refer to Appendix~\ref{appendix:rk} for the proof.
\endIEEEproof
\end{Lem}

The total power consumption of the above system consists of two parts, e.g., the power consumption of the BS, $P_{BS}$, and the power consumption of the RIS, $P_{RIS}$. According to the power consumption model as defined in \cite{long2021active}, $P_{BS}$ and $P_{RIS}$ are given by,
\begin{align}
    P_{BS}(\mathbf{W}) & = \mu_{BS} \sum_{k=1}^{K}||\mathbf w_k||^2 + P^{S}_{BS}, \\
    P_{RIS}(\mathbf{\Theta}, \mathbf{W}) & = \mu_{RIS} \left(\sum_{k=1}^{K}||\mathbf\Theta\mathbf{H}_{br} \mathbf w_k||^2+\sigma_{R}^2||\mathbf\Theta||^2_{F}\right) \nonumber \\
    & + P^{S}_{RIS}+N_{RIS}P_{u}, 
\end{align}
where $P^{S}_{BS}$ , $P^{S}_{RIS}$ and $P_{u}$ denote the static power consumption at the base station ,  at the active RIS, and per element of active RIS respectively.  $\mu_{BS}$  and $\mu_{RIS}$  represent the power amplification coefficients at the base station and RIS, respectively. 

The following assumptions are adopted through the rest of this paper. First, to ensure the service quality of each user, a minimum average transmission rate, $C_k^{\min}$, is guaranteed, e.g., $\bar{R}_k\geq C^{\min}_{k}, \forall k$. Second, the total emitted power for BS and RIS is limited, such that $\sum_{k=1}^{K}||\mathbf w_k||^2\leq P_{BS}^{\max}$ and $\sum_{n=1}^{N_{RIS}}||\boldsymbol\Theta\mathbf{H}_{br} \mathbf w_k||^2+\sigma_{R}^2||\boldsymbol\Theta||^2_{F}\leq P_{RIS}^{\max}$. Last but not least, due to the hardware limitation of RIS components \cite{wang2024reconfigurable}, the reflection coefficients $\{\theta_n\}$ are selected from a discrete set $\Omega_{\theta}$ with size $2^{L}$, e.g., $\theta_n \in \Omega_{\theta}$ and $|\Omega_{\theta}| = 2^{L}$.
\subsection{Problem Formulation}
Instead of using the conventional energy efficiency metric, we adopt the network IREE as defined in \cite{yu2022novel}, and summarize the main results in the following definition.

\begin{Def}[Network IREE \cite{yu2022novel}] The network IREE of the active RIS enabled system based on outdated CSI ,  $\eta_{IREE}\left(\mathbf{\Theta}, \mathbf{W} \right)$,  is defined as,
\begin{eqnarray}
\label{eq:IEE}
    \eta_{IREE}\left(\mathbf{\Theta}, \mathbf{W} \right) & = & \frac{\sum_{k=1}^{K}\min \{\bar{R}_k\left(\mathbf{\Theta}, \mathbf{W}\right), D_k\}}{P_{BS}(\mathbf{W}) + P_{RIS}(\mathbf{\Theta}, \mathbf{W})},
\end{eqnarray}
where $D_k$ is the throughput requirement of the $k$-th user\footnote{In practical systems, the user demand is collected on a per-second basis and assumed to be constant during the transmission period as pointed by \cite{azari2021energy}.}. 
\label{def:iree}
\end{Def}

Based on Definition~\ref{def:iree}, we can maximize the network IREE by tuning the beamforming metrices $\boldsymbol \Theta$ and $\mathbf W$ via the following optimization problem.

\begin{Prob}[Original Problem] 
\label{prob:IREE} 
The network IREE of RIS enabled 6G network can be maximized by solving the following problem.  
\begin{eqnarray}
    \underset{(\boldsymbol \Theta, \mathbf W)}{{\textrm{maximize}}} && {\eta} _{IREE} (\boldsymbol \Theta, \mathbf W)\label{eq:IREE}\\
    \textrm{subject to}&& \sum_{k=1}^{K}||\mathbf w_k||^2\leq P_{BS}^{\max} ,\label{constrain:max_power_bs}\\&&  \sum_{n=1}^{N_{RIS}}||\boldsymbol\Theta\mathbf{H}_{br} \mathbf w_k||^2+\sigma_{R}^2||\boldsymbol\Theta||^2_{F}\leq P_{RIS}^{\max},\label{constrain:max_power_ris} \\
     && \bar{R}_k(\boldsymbol \Theta, \mathbf W) \geq C^{\min}_{k}, \quad \forall k \in [1,K],\label{constrain:capability}\\
    && \theta_{n} \in \Omega_{\theta}, \quad \forall n \in [1,N_{RIS}]. \label{constrain:positivq}
\end{eqnarray}
\end{Prob} 

In general, Problem \ref{prob:IREE} is difficult to solve using conventional methods due to the following reasons. On one hand, since the reflection coefficients $\{\theta_n\}$ are selected from a discrete set, the above problem is a typical non-convex combinatorial optimization problem. On the other hand, the fractional programming structure of the IREE metric, coupled with the entanglement between $\boldsymbol \Theta$ and $\mathbf W$, makes the problem even more challenging to handle.

\section{Proposed IREE Maximization Scheme} \label{sect:proposed_scheme}
In this section, we transform the aforementioned
non-convex combinatorial optimization problem using the quadratic transform \cite{shen2018fractional} and relaxation methods, and we subsequently propose our IREE maximization scheme , Alternating Optimization Successive Approximation (AOSO)  algorithm.

In order to deal with the fractional programming structure, we adopt the Quadratic Transform\cite{shen2018fractional} technique by introducing an auxiliary variable, $y$, and transform the original problem into the following structure.

\begin{Prob}[Quadratic Transformed Problem] 
\label{prob:IREE2} With some mathematical manipulations as summarized in Appendix~\ref{app:1to2}, the equivalent quadratic transformed IREE maximization problem is given as follows. 
\begin{eqnarray}
\label{g_w}
    \underset{ (\mathbf W ,  \mathbf \Theta, y)
   } {\textrm{maximize}} && f(\mathbf W,\boldsymbol \Theta, y)= 2y\sqrt{\sum_{k=1}^{K}\min \{ \bar{R}_k\left(\mathbf{\Theta}, \mathbf{W}\right), D_k \} } \nonumber \\
   && -y^2 \left( P_{BS}\left(\mathbf{W}\right)+P_{RIS}\left(\mathbf{\Theta}, \mathbf{W}\right)\right), \\
    \textrm{subject to} &&\eqref{constrain:max_power_bs}-\eqref{constrain:positivq}. \notag 
\end{eqnarray}
\end{Prob}

However, Problem \ref{prob:IREE2} remains non-concave and complex due to the non-convexity of $ f(\mathbf{W}, \boldsymbol{\Theta}, y) $ and the constraints \eqref{constrain:capability} and \eqref{constrain:positivq}. Therefore, we first equivalently transform $ f(\mathbf{W}, \boldsymbol{\Theta}, y) $ into $ g(\mathbf{W}, \boldsymbol{\Theta}, \mathbf{u}) $ by introducing the variable $ \mathbf{u} = (u_1, \dots, u_k, \dots, u_K) $:
\begin{align}
    &g(\mathbf{W}, \boldsymbol{\Theta}, y, \mathbf{u}) = 2y\sqrt{\sum_{k=1}^{K} u_k} - y^2(P_{BS} + P_{RIS}) \notag \\
    & u_k \leq \bar{R}_k, \quad \forall k \in [1,K], \label{eq:mu_bs_r} \\
    &u_k \leq D_k, \quad \forall k \in [1,K], \label{eq:mu_bs_d} \\
    &u_k \geq C_k^{\min}, \quad \forall k \in [1,K], \label{eq:qd_cap}
\end{align}
where constraints \eqref{eq:mu_bs_r} and \eqref{eq:mu_bs_d} hold with equality for the optimal solution, and \eqref{eq:qd_cap} is a relaxation of \eqref{constrain:capability}.
Next, to address the complex constraint \eqref{eq:mu_bs_r}, we introduce a slack variable $ \mathbf{v} = (v_1, \dots, v_k, \dots, v_K) $. The constraint \eqref{eq:mu_bs_r} can then be rewritten as:
\begin{align}
    u_k &\leq \log_2\left(1 + \frac{|\rho \hat{\mathbf{h}}_{ru,k} \boldsymbol{\Theta} \mathbf{H}_{br} \mathbf{w}_{k}|^2}{v_k}\right),  \forall k \in [1,K], \label{eq:ck_bs} \\
   v_k &\geq \sum_{i=1}^{K} ||\mathbf{Q}_{ki} \boldsymbol{\Theta} \mathbf{H}_{br} \mathbf{w}_{i}||^2  + ||\mathbf{Q}_{k0} \boldsymbol{\Theta}||^2 \sigma_{R}^2 + \sigma_{n}^2,\forall k \in [1,K] \label{eq:v_bs}
\end{align}
where \eqref{eq:v_bs} ensures the equivalence of the transformation.
We further process \eqref{eq:ck_bs} by applying the quadratic transform once again and introduce a slack variable $ \mathbf{z} = (z_1, \ldots, z_k, \ldots, z_K) $, recasting \eqref{eq:ck_bs} as:
\begin{align}
    \label{eq:c_bs_qd}
    u_k \leq \log_2\left(1 + 2 z_k \textbf{Re}(\mathbf{H}_k \mathbf{w}_k) - z_k^H v_k z_k\right), \quad \forall k \in [1,K],
\end{align}
where $ \mathbf{H}_k = \rho \hat{\mathbf{h}}_{ru,k} \boldsymbol{\Theta} \mathbf{H}_{br} $.
After the above discussion, we reformulate Problem \ref{prob:IREE2} into the following problem:
\begin{Prob} 
We can solve the following optimization problem by iteratively optimizing $ \mathbf{z} $ and the other variables. The proof of the transformation from Problem \ref{prob:IREE2} to Problem \ref{prob:IREE3} can be referenced in Appendix \ref{app:pro3}.
\label{prob:IREE3}
\begin{eqnarray}
    \underset{ (\mathbf{W}, \boldsymbol{\Theta}, y, \mathbf{u}, \mathbf{v}, \mathbf{z}) } {\textrm{maximize}} &&  g(\mathbf{W}, \boldsymbol{\Theta}, y, \mathbf{u}, \mathbf{v}, \mathbf{z}), \\
    \textrm{subject to} && \eqref{constrain:max_power_bs}, \eqref{constrain:max_power_ris}, \eqref{constrain:positivq}, \eqref{eq:mu_bs_r}-\eqref{eq:qd_cap}, \eqref{eq:c_bs_qd}. \notag 
\end{eqnarray}
\end{Prob}
By addressing Problem \ref{prob:IREE3}, we can achieve the objective of maximizing IREE. Although Problem \ref{prob:IREE3} is still not a concave problem, it can be quickly solved by decoupling it into two subproblems—optimizing $ \mathbf{W} $ and $ \boldsymbol{\Theta} $ separately—following the approach of alternating optimization.

Based on this, we transform Problem \ref{prob:IREE3} into two subproblems, \ref{prob:sub1} and \ref{prob:sub2}. By alternately optimizing these two subproblems, we can eventually obtain the solution to Problem \ref{prob:IREE3}, thereby maximizing the IREE.

\begin{subProb}
When $\boldsymbol\Theta$ is fixed, the optimization of other variables can be formulated as the following optimization problem.
\label{prob:sub1}
\begin{eqnarray}
    \underset{ (\mathbf W,y,\mathbf u,\mathbf v,\mathbf z  ) 
   } {\textrm{maximize}} &&  g(\mathbf W,y,\mathbf u,\mathbf v,\mathbf z),\\
    \textrm{subject to} &&\eqref{constrain:max_power_bs},\eqref{constrain:max_power_ris},\eqref{eq:mu_bs_d},\eqref{eq:qd_cap},\eqref{eq:v_bs},\eqref{eq:c_bs_qd}.\notag \\
\end{eqnarray}
\end{subProb}
\begin{subProb}
When $\mathbf W$ is fixed, the optimization of other variables can be formulated as the following optimization problem.
\label{prob:sub2}
\begin{eqnarray}
    \underset{ ( \boldsymbol\Theta,y,\mathbf u,\mathbf v,\mathbf z  ) 
   } {\textrm{maximize}} &&  g(\boldsymbol\Theta,y,\mathbf u,\mathbf v,\mathbf z),\\
    \textrm{subject to} &&\eqref{constrain:max_power_ris},\eqref{constrain:positivq},\eqref{eq:mu_bs_d},\eqref{eq:qd_cap},\eqref{eq:v_bs},\eqref{eq:c_bs_qd}.\notag \\
\end{eqnarray}
\end{subProb}
Next, we alternately optimize $\mathbf{W} $ and $ \boldsymbol{\Theta} $ by solving these two subproblems separately.

\subsection{\texorpdfstring{Alternate optimization of $\mathbf{W}$}{Alternate optimization W}}

We optimize $ \mathbf{W} $ while keeping $ \boldsymbol{\Theta} $ fixed. First, we fix all variables except for $ y $ and solve Problem \ref{prob:sub1} to obtain the optimal value $ y^* $:
\begin{equation}
    y^* = \frac{\sqrt{\sum_{k=1}^{K} u_k}}{\sum_{k=1}^{K} ||\mathbf{w}_k||^2 + \sum_{n=1}^{N_{RIS}} ||\boldsymbol{\Theta} \mathbf{H}_{br} \mathbf{w}_k||^2 + \sigma_{R}^2 ||\boldsymbol{\Theta}||_{F}^2}.
    \label{eq:y_bs}
\end{equation}
Next, we fix all variables except for $ \mathbf{z} $ and solve Problem \ref{prob:sub1} to obtain the optimal value $ \mathbf{z}^* $:
\begin{equation}
    \mathbf{z}_k^* = \frac{\textbf{Re}(\mathbf{H}_k \mathbf{w}_k)}{v_k} \quad \forall k \in [1, K].
    \label{eq:z}
\end{equation}
At this point, with $ y $, $ \mathbf{z} $, and $ \boldsymbol{\Theta} $ fixed, Problem \ref{prob:sub1} becomes a concave problem, which can be solved using toolboxes such as CVXPY \cite{cvxpy}. 
\subsection{\texorpdfstring{Alternate Optimization of $ \boldsymbol{\Theta} $}{Alternate Optimization of Theta}}

We optimize $ \boldsymbol{\Theta} $ while keeping $ \mathbf{W} $ fixed. First, we relax the discrete values in constraint \eqref{constrain:positivq} to continuous values, i.e., $ |\theta_{n}| \leq \alpha_{\max} $. We define $ \mathbf{T}_{ki}^{'} = \mathbf{Q}_{ki} \textbf{diag}(\mathbf{H}_{br} \mathbf{w}_i) $ and $ \mathbf{T}_k = \rho \hat{\mathbf{h}}_{ru,k} $. Problem \ref{prob:sub2} can then be reformulated as follows:
\begin{Prob} 
\label{prob:IREE4}
\begin{eqnarray}
    \underset{ \{\boldsymbol{\Theta},y,\mathbf u,\mathbf v,\mathbf z  \} 
   } {\textrm{maximize}} &&  g(\boldsymbol{\Theta},y,\mathbf u,\mathbf v,\mathbf z ),\\
    \textrm{subject to} &&\eqref{constrain:max_power_ris},\eqref{eq:mu_bs_d},\eqref{eq:qd_cap},\notag \\
   &&  v_k\geq \sum_{k=1}^K|| \mathbf{T}_{ki}^{'}\boldsymbol \theta ||^2+|| \mathbf Q_{ki} \boldsymbol  \Theta||^2\sigma_{R}^2\notag\\
   &&\quad\quad+\sigma_n^2,\quad\forall k \in [1,K]\label{eq:v_ris}\\
   && u_k\leq\log_2(1+2 z_k\textbf{Re}(\mathbf T_k \boldsymbol \theta_k)\notag\\
   &&\quad\quad- z^H_k{v_k} z_k),\quad\forall k \in [1,K]\label{eq:ris_z}\\
   &&|\theta_{n}|\leq \alpha_{\max}\quad \forall n \in [1,N_{RIS}].\label{eq:continus}
\end{eqnarray}
\end{Prob}
Next, we fix all variables except for $y $ and solve Problem \ref{prob:IREE4} to obtain the optimal value $y^* $:
\begin{equation}
    y^* = \frac{\sqrt{\sum_{k=1}^{K} u_k}}{\sum_{k=1}^{K} ||\mathbf{w}_k||^2 + \sum_{n=1}^{N_{RIS}} ||\boldsymbol{\Theta} \mathbf{H}_{br} \mathbf{w}_k||^2 + \sigma_{R}^2 ||\boldsymbol{\Theta}||_{F}^2}.
    \label{eq:y_ris}
\end{equation}
We then fix all variables except for $\mathbf{z} $ and solve Problem \ref{prob:IREE4} to obtain the optimal value $\mathbf{z}^* $:
\begin{equation}
\label{eq:z_ris}
    \mathbf{z}_k^* = \frac{\textbf{Re}(\mathbf{T}_k \boldsymbol{\theta})}{v_k} \quad \forall k \in [1,K].
\end{equation}
At this point, with $y $, $\mathbf{z} $, and $\mathbf{W} $ fixed, Problem \ref{prob:IREE4} becomes a concave problem, which can be solved using toolboxes such as CVXPY.

Finally, we use a successive approximation method to map $\boldsymbol{\theta} $ from the continuous domain back to the discrete domain, thereby achieving the goal of solving Problem \ref{prob:sub2}. The proposed method begins by solving Problem \ref{prob:IREE4} to obtain $ \boldsymbol{\Theta}^* $. This solution is subsequently mapped to the discrete domain using the following update rule, yielding the rounded values $ \boldsymbol{\Theta}^*_r $:
\begin{align}
    \label{eq:get_discret}
    \theta^*_{r,n} = \text{argmax}_{\theta \in \mathbb{S}_{\theta}} |\theta_{r,n} - \theta| \quad \forall n \in [1, N_{RIS}].
\end{align}
And we calculate the distance $ d_n $ between $ \theta^*_{r,n} $ and its pre-mapping continuous value $ \theta^*_n $, normalizing it as follows:
\begin{equation}
    \label{eq:d}
    d_n = \frac{|\theta^*_{r,n} - \theta^*_n|}{|\theta^*_n|} \quad \forall n \in [1, N_{RIS}].
\end{equation}

From these distances, we select the half of $ \boldsymbol{\Theta}^* $ with the smallest $ d $ values to keep fixed for the next iteration. We then resolve Problem \ref{prob:IREE4} to obtain an updated $ \boldsymbol{\Theta}^* $. In subsequent iterations, we continue this process, selecting the half of $ \boldsymbol{\Theta}^* $ with the smallest distance among those not yet fixed, and maintaining them unchanged. This iterative process continues until the predetermined number of iterations $ Q $ is reached. The corresponding algorithm is outlined in Algorithm \ref{alg:alter_dis}.
 \begin{algorithm}[t]
\caption{  successive approximation method}
 \label{alg:alter_dis}  
\SetKwInOut{Input}{input}
\SetKwInOut{Output}{output}
\SetKwRepeat{for}{Do}{if}
Input: $\boldsymbol \Theta_r^{(0)}$,$\boldsymbol \Theta^{best}=\boldsymbol \Theta_r^{(0)}$,$\$P=\varnothing$,$y$, $\mathbf S$; 

\For{$i=1,2,\ldots,Q$} 
{

{under the given constraint $\theta_p = \theta_{r,p},p\in P$
, solve the concave maximization problem: ${\textrm{max}}\quad g(\boldsymbol{\Theta},  y,\mathbf u,\mathbf v,\mathbf  z)$ and update $\boldsymbol \Theta^{(i)}$};

update $ \boldsymbol \Theta_r^{(i)}$ by
\eqref{eq:get_discret};

{calculate the distance $\{d_n\}_{n=1}^{N_{RIS}}$ and select the indexs of the smallest half to be placed into the set $P$
 };
 
\If {$|g(\boldsymbol{\Theta}_r^{i},  y,\mathbf u^{i},\mathbf v^{i},\mathbf z)>g(\boldsymbol{\Theta}_r^{i-1},  y,\mathbf u^{i-1},\mathbf v^{i-1},\mathbf z)$}
{
$\boldsymbol \Theta^{best}=\boldsymbol \Theta_r^{(i)}$;

}

{i=i+1};
}
\textbf{Output:}$\boldsymbol\Theta=\boldsymbol \Theta^{best}$

\end{algorithm}

Based on the above derivation, we designed Algorithm \ref{alg:alter_iree} (AOSO) to address problem \ref{prob:IREE}  .
 \begin{algorithm}[t] 
\caption{ Alternating optimization $\eta_{IREE}$ Maximization Scheme} 
\label{alg:alter_iree} 
\SetKwInOut{Input}{input}
\SetKwInOut{Output}{output}
\SetKwRepeat{for}{Do}{if}
Initialization: $\mathbf W^{(0)},\boldsymbol \Theta^{(0)}$; 

\For{$i=1,2,\ldots$} 
{
update $  y^{(2i-1)},\mathbf z^{(2i-1)}$ by
\eqref{eq:y_ris},\eqref{eq:z};

{Given $\mathbf W^{(i-1)}$, use Algorithm \ref{alg:alter_dis} to  update $\boldsymbol \Theta^{(i)}$};

update $  y^{(2i)},\mathbf z^{(2i)}$ by
\eqref{eq:y_bs},\eqref{eq:z_ris};

{Given $\boldsymbol \Theta^{(i)}$,solve the concave maximization problem:
${\textrm{max}}\quad g(\mathbf W, y^{(2i)},\mathbf u,\mathbf v,\mathbf z^{(2i)})$ and update $\mathbf W^{(i)}$};

update $\eta_{\text{IREE}}^{i-1}$ by \eqref{eq:IEE}

\If {$|\eta_{\text{IREE}}^{i}-\eta_{\text{IREE}}^{i-1}|<\epsilon$}
{

\textbf{Output:}$\mathbf W, \mathbf \Theta$
}
}
\end{algorithm}
\section{Performance Analysis}
In this section, we analyze the convergence and complexity of the proposed algorithm.
\subsection{Analysis of Convergence }
Based on the proposed algorithm and the problem derivation process, we have the following theorem:
\begin{Thm}
    Algorithm \ref{alg:alter_iree} (AOSO) produces a convergent sequence
of the objective values of \eqref{eq:IREE}.
\IEEEproof 
\label{proof:converge}
Please refer to Appendix \ref{app:converge}
\endIEEEproof
\end{Thm}
\begin{figure}[!t]
       
   \centering
    \includegraphics[width=1.0\columnwidth]{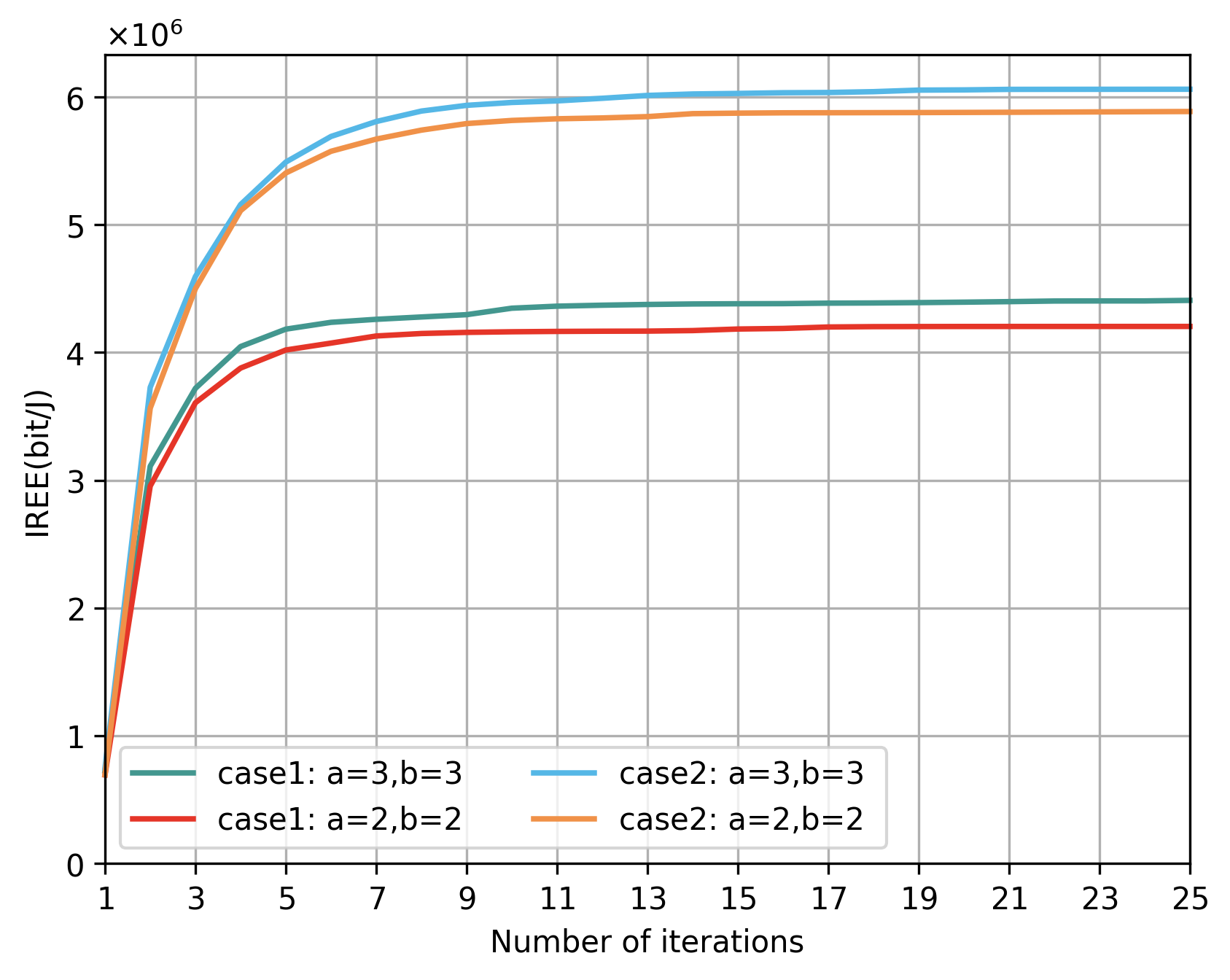}
    \caption{Convergence of Algorithm \ref{alg:alter_iree}}
        \label{fig:converge}
    \end{figure}
    
The convergence of our proposed algorithm is illustrated in Fig.\ref{fig:converge}. The entire experiment is conducted under the condition of \( P_{\text{all}}^{\max} = 30 \) dBm. The specific experimental setup is the same as in the following Section \ref{sect:num_res}. It can be observed that, regardless of whether it is Case 1 or Case 2, the algorithm converges rapidly within 20 iterations. Furthermore, the quantization levels do not affect the convergence speed of the algorithm. This demonstrates that our proposed algorithm achieves convergence in a relatively short number of iterations and exhibits robust performance.
\subsection{Analysis of Complexity }
For Algorithm \ref{alg:alter_iree}, the algorithm complexity is related to the iteration speed and the complexity of solving convex problems. According to \cite{lobo1998applications}, we know that the complexity of a second-order cone programming (SOCP) problem is $ O(x_1^2x_2) $, where $ x_1 $ is the number of variables and $ x_2 $ is the number of constraints. Problems \ref{prob:sub1} and \ref{prob:IREE4} are categorized as SOCP problems. For Problem \ref{prob:sub1}, it has $(N_t + 3)K$ variables and $5K+2$ constraints, so its corresponding complexity is:
\begin{eqnarray}
    \label{eq:o1}
    \mathcal{O}_1\left((N_tK + 3K)^2(5K+2)\right).
\end{eqnarray}
For Problem \ref{prob:IREE4}, it has $N_{RIS} + 3K$ variables and $5K+N_{RIS}+1$ constraints, so its corresponding complexity is:
\begin{eqnarray}
    \label{eq:o2}
    \mathcal{O}_2\left((N_{RIS} + 3K)^2(5K+N_{RIS}+1)\right).
\end{eqnarray}
Assuming convergence after $I$ iterations and that solving convex optimization problems has polynomial-level complexity, and the iteration updates for variables such as $y,\mathbf{Z},\mathbf{S}$ can be neglected compared to the complexity of convex optimization problems. Additionally, in the successive approximation sub-algorithm, Problem \ref{prob:IREE4} is solved $ Q $ times, and the sorting algorithm is performed $ Q $ times as well.
The complexity of Algorithm \ref{alg:alter_iree} can be estimated as
\begin{align}
    \label{eq:comlex}
    &\mathcal{O}\left(I(\mathcal{O}_1 + Q\mathcal{O}_2 + QN_{RIS}\log_2(1+N_{RIS}))\right).
\end{align}
Through our numerical experiments below, the algorithm quickly converges within a few iterations, so the overall complexity is related to the number of transmit antennas $N_t$, the number of users $K$, and the RIS element count $N_{RIS}$.

\section{Numerical Results } 
\label{sect:num_res}
 In this section, numerical examples are provided to validate.
In our simulation setup, the system consists of $N_t$ transmit antennas, an active RIS wtih $N_{max}$ elements, and $K$ single-antenna users. And the base station is located at the origin, with the active RIS deployed on the x-axis at a distance of $10$m  from the base station. $K$ users are evenly distributed along a quarter-circle with a radius of $80$m, centered at the base station. Additionally, due to the low penetration capability of millimeter waves and the impact of obstacles, the direct link between the base station and the users is blocked.

  The channel parameters for the links $\boldsymbol{h}^H_{ru,k}$ and $\boldsymbol{H}_{br}$ are configured according to the Saleh-Valenzuela (SV) channel model \cite{saleh1987statistical}, which is a typical millimeter-wave communication model. Their corresponding channels are modeled as follows:
\begin{align}
&\mathbf{H}_{br} = \sqrt{\frac{N_tN_{RIS}}{L_{br,k}+1}} \sum_{i=0}^{L_{br}} \xi_t \xi_{R} h_{br}^{i} \mathbf{a}(\theta_{br}^{i}, \phi_{br}^{i}) \mathbf{b}^H(\theta_{br}^{i}, \phi_{br}^{i}),\label{eq:channelbu}\\   
&\mathbf{h}_{ru,k} = \sqrt{\frac{N_{RIS}}{L_{ru,k}+1}} \sum_{i=0}^{L_{ru,k}} \xi_r h_{ru,k}^{i} \mathbf{b}(\theta_{ru,k}^{i}, \phi_{ru,k}^{i}),
\label{eq:channelru}
\end{align}
where $ \xi_t $, $ \xi_R $, and $ \xi_r $ denote the antenna gains at the transmitter, RIS, and receiver, respectively. $ i=0 $ represents the line-of-sight (LOS) component, while $ i \neq 0 $ represents the non-line-of-sight (NLOS) component. The channel fading coefficient
 $h_{br}^{i}$ and $h_{ru,k}^{i}$ follow $h_{br}^{i}, h_{ru,k}^{i} \sim \mathcal{CN}(0,10^{-PL})$, where $PL$ represents the path loss. $L_{br}$ and $L_{ru,k}$ denote the number of scattering paths between the base station and the RIS, and between the $k$-th user and the RIS, respectively. We set $L_{br} = 4$ and $L_{ru,k} = 4$. $\boldsymbol{a}(\theta,\phi)$ represents the steering vector of the uniform linear array (ULA), and $\boldsymbol{b}(\theta,\phi)$ represents the steering vector of the uniform planar array (UPA). Here, $\theta$ and $\phi$ denote the azimuth and elevation angles of departure or arrival (AOA or AOD), respectively. The path loss is defined as $PL = 32.4 \text{dB} + 20\log_{10}f_c + 10\alpha \log_{10}d + \xi$, where $\alpha$ is the path loss exponent, $d$ is the distance of the link, and $\xi \sim \mathcal{CN}(0, \sigma_{\xi}^2)$ represents the shadowing exponent. In LOS and NLOS scenarios, $\alpha = 2$ and $\alpha = 2.82$, with $\xi \sim \mathcal{CN}(0,\sigma_{\xi}^2=10^{0.4}\text{dB})$ for LOS and $\xi \sim \mathcal{CN}(0,\sigma_{\xi}^2=10^{0.82}\text{dB})$ for NLOS, as described in \cite{zhou2022fairness}\cite{generation2019technical}. 

The power consumption per unit element of the active RIS is $ P_u = P_{DC} + P_r $, while for the passive RIS, it is $ P_u = P_r $, where $ P_r $ is 1.5 mW, 3 mW, and 4.5 mW for 1-bit, 2-bit, and 3-bit phase quantization levels, respectively \cite{wang2024reconfigurable}. Additionally, the amplifier's DC power consumption at the active RIS is 20.5 mW \cite{gao201924}.We set the quantization levels for amplitude and phase to be $a=2$ and $b=2$, respectively.
We set the total system power budget as $ P_{all}^{\max} = P_{BS}^{\max} + P_{RIS}^{\max} $, with $ P_{BS}^{\max} = 0.99 P_{all}^{\max} $ and $ P_{RIS}^{\max} = 0.01 P_{all}^{\max} $.
Each user's minimum requirement $ C_{\text{min},k} $ is set to $ 0.1D_{k} $, capped at a maximum of 1 Mbit/s. The traffic demand for each user $ D_k $ is sampled from a log-normal distribution \cite{wang2015approach}. We consider two different traffic scenarios:
\begin{itemize}
\item Case 1: High average traffic with a variance of 10 Mbit/s and a mean of 60 Mbit/s.
\item Case 2: Low average traffic with a variance of 10 Mbit/s and a mean of 100 Mbit/s.
\end{itemize}
The overall experimental results are obtained by averaging multiple Monte Carlo samples. The remaining system parameters are listed in Table \ref{tb:args} \cite{huang2019reconfigurable} \cite{zhao2021joint}.
\begin{table}[h]
\caption{Parameter Settings}
\label{tb:args}
\begin{tabular}{|c|c|}
\hline
\textbf{Parameter} & \textbf{Value} \\
\hline

Number of Antennas $N_t$& $12$\\
Number of Users $K$& $8$\\
 Number of RIS elements $N_{RIS}$& $20$ \\
Static power consumption of BS $P_{\text{BS}}^S$ & $9$dBW \\
Static power consumption of RIS $P_{\text{RIS}}^S$ & $1.5$W\cite{wang2024reconfigurable}\\
   Power amplification coefficients at BS  $\mu_{BS}$ & $1.2$\\
    Power amplification coefficients at  RIS   $\mu_{RIS}$ & $1.2$\\
    correlation coefficient $\rho$ & $0.9$\\
   Amplification factor of the amplifier $\alpha_{max}$ & $23$dB\cite{gao201924}\\
Noise power $\sigma^2_n,\sigma^2_R$ & $-94$dBm\\
Bandwidth $BW$ & $100$MHZ\\
carrier frequency of millimeter wave $f_c$ & $28$GHZ \\
 Algorithm convergence factor $ \epsilon$& $0.001$\\
 Successive approximation iterations  $ Q$& $3$\\
 the antenna gains of RIS $\xi_{ris}$& $19.8$dBi\cite{tang2023transmissive}\\
Transmit
antenna gains$\xi_t$& $12.98$dBi\cite{amith2024corporate}\\
Receive
antenna gains$\xi_r$& $5.51$dBi\cite{hwang2019quasi}\\
\hline
\end{tabular}
\end{table}

\subsection{Algorithm performance comparison}

In this subsection, we present simulation results to evaluate the performance of our proposed algorithm in solving the IREE maximization problem for RIS-assisted millimeter-wave communication systems. We compare our proposed scheme with other schemes. The other schemes are as follows:
\begin{itemize}
\item continuous condition: Assumes that the beamforming matrix at the RIS can be continuously adjusted.In this case, $\boldsymbol{\Theta}$ is directly obtained in each iteration of our algorithm by solving the problem ${\textrm{max}}\quad g(\boldsymbol{\Theta},  y,\mathbf u,\mathbf v,\mathbf z)$, without the need for the successive approximation algorithm. 
\item EE Baseline: Using the method from \cite{niu2023active} under the same experimental conditions, the corresponding $\eta_{IREE}$ is then calculated. 
\item Passive RIS: Applies our method to a passive RIS, where the number of elements is set to $ N_{{max}} = 100 $ to ensure the same power consumption as the active RIS.
\end{itemize}
\begin{figure*}[!t]
        \centering
        \subfloat[]{
        \includegraphics[width=0.43\textwidth]{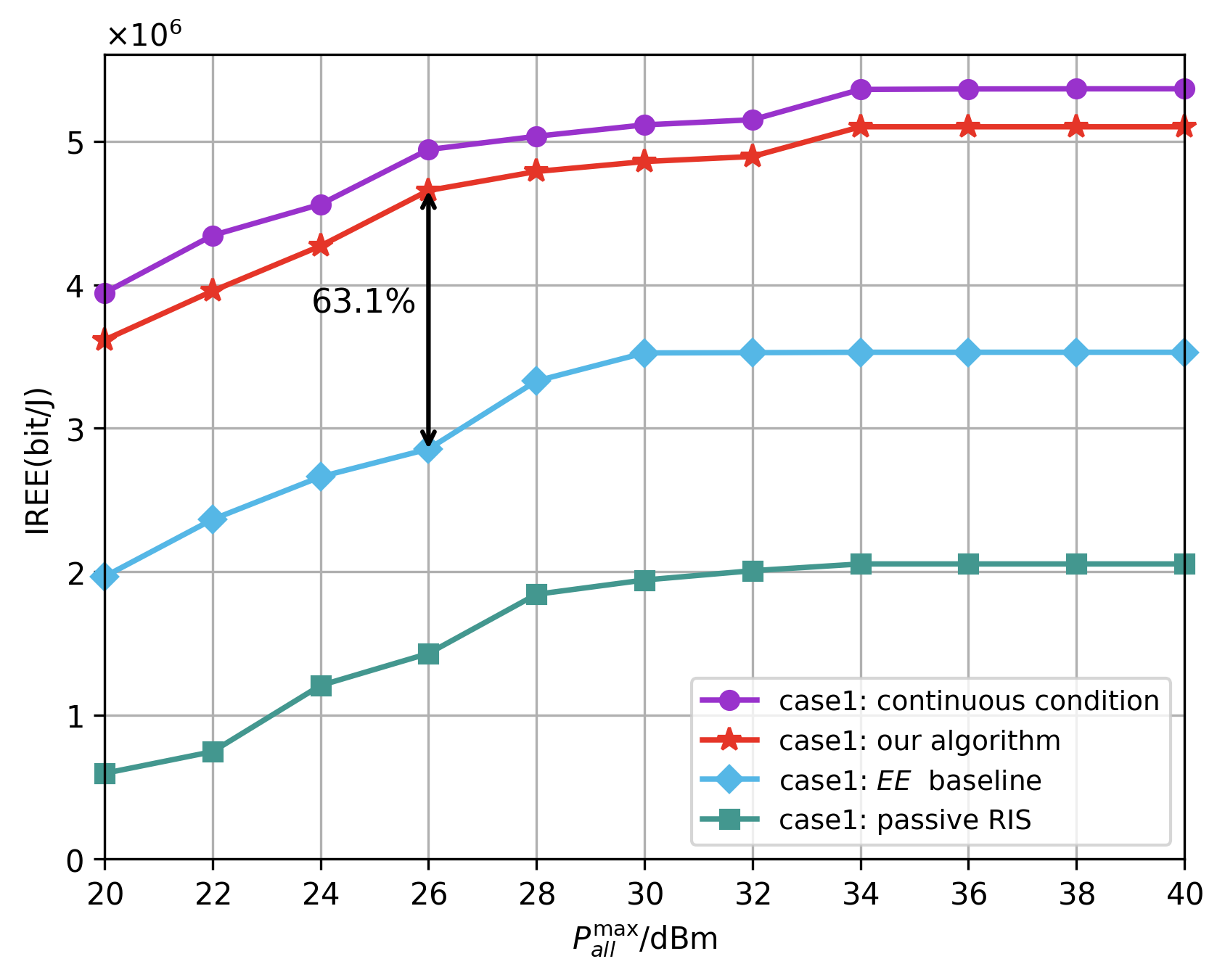} 
    }
    \hfill
    \subfloat[]{
        \includegraphics[width=0.43\textwidth]{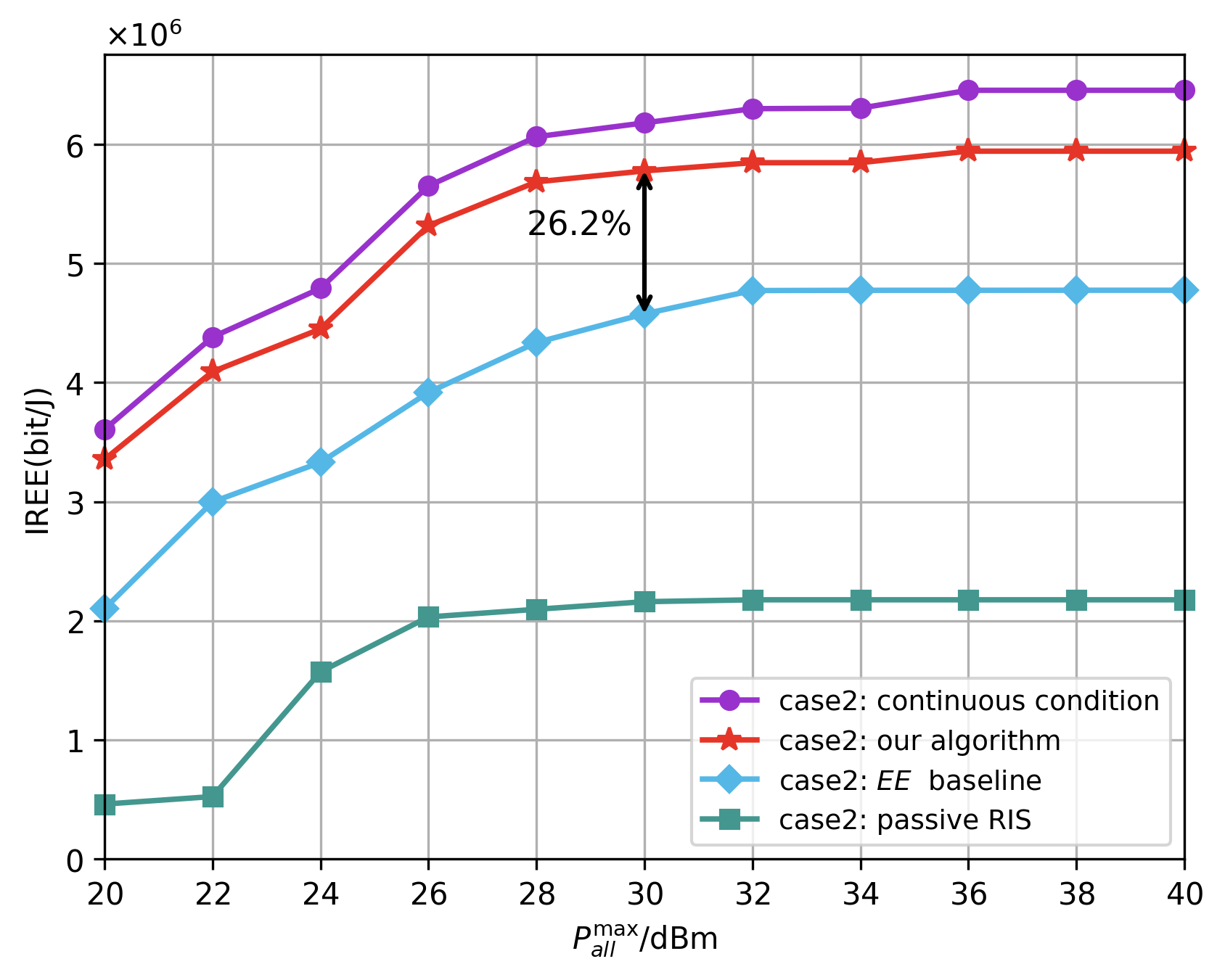} 
    }
    \caption{$\eta_{IREE}$ versus $P_{max}$ in different  algorithms}
        \label{fig:dif_alg}
    \end{figure*}
The specific performance comparison is shown in Fig. \ref{fig:dif_alg}. The experimental results indicate that the $\eta_{IREE}$ increases with the rise in $P^{{\max}}_{all}$ until it reaches a certain threshold. After $P^{{\max}}_{all}$ hits this threshold, the $\eta_{IREE}$ becomes constant. This is mainly because, as $P^{{\max}}_{all}$ increases, SINR at the receiver gradually improves, leading to enhanced user capacity and better system performance. However, due to the constraints on each user's actual traffic demand, the improvement in IREE performance reaches a corresponding upper limit. 

First, when comparing to the passive RIS baseline, our method significantly outperforms it in both Case 1 and Case 2. This is mainly because the active RIS can amplify the received signal while consuming relatively little active power, unlike the passive RIS which can only modify the phase of the received signal. This allows the active RIS system to significantly enhance the received SNR for each user, thereby effectively improving the overall system performance. Furthermore, our approach demonstrates considerable improvement over the EE baseline, with enhancements of $63.1\%$ and $40.5\%$ in Case 1 and Case 2, respectively. This is because the algorithm designed for the traditional EE metric fails to effectively match user demand and traffic, resulting in lower IREE performance. Our method effectively accounts for the matching between user demand and the provided capacity, thereby significantly improving the IREE performance. Additionally, since the user demands are higher in Case 2 compared to Case 1, instances where the actual capacity exceeds user demand are reduced. As a result, the gap between the EE baseline method and our approach is somewhat diminished.
Lastly, compared to the continuous case, our algorithm's overall performance is quite close. Since our proposed algorithm, in each iteration of $\boldsymbol{\Theta}$, approaches the optimal value multiple times from the discrete domain. In cases with low quantization levels, it can achieve results close to those in the ideal scenario . 
The above results demonstrate that our proposed algorithm effectively maximizes the IREE metric, while the use of active RIS significantly enhances system performance compared to passive RIS.
\subsection{The impact of Quantization Levels on the Reflection Coefficients at RIS}
\begin{figure*}
        \centering
       \subfloat[]{
        \includegraphics[width=0.43\textwidth]{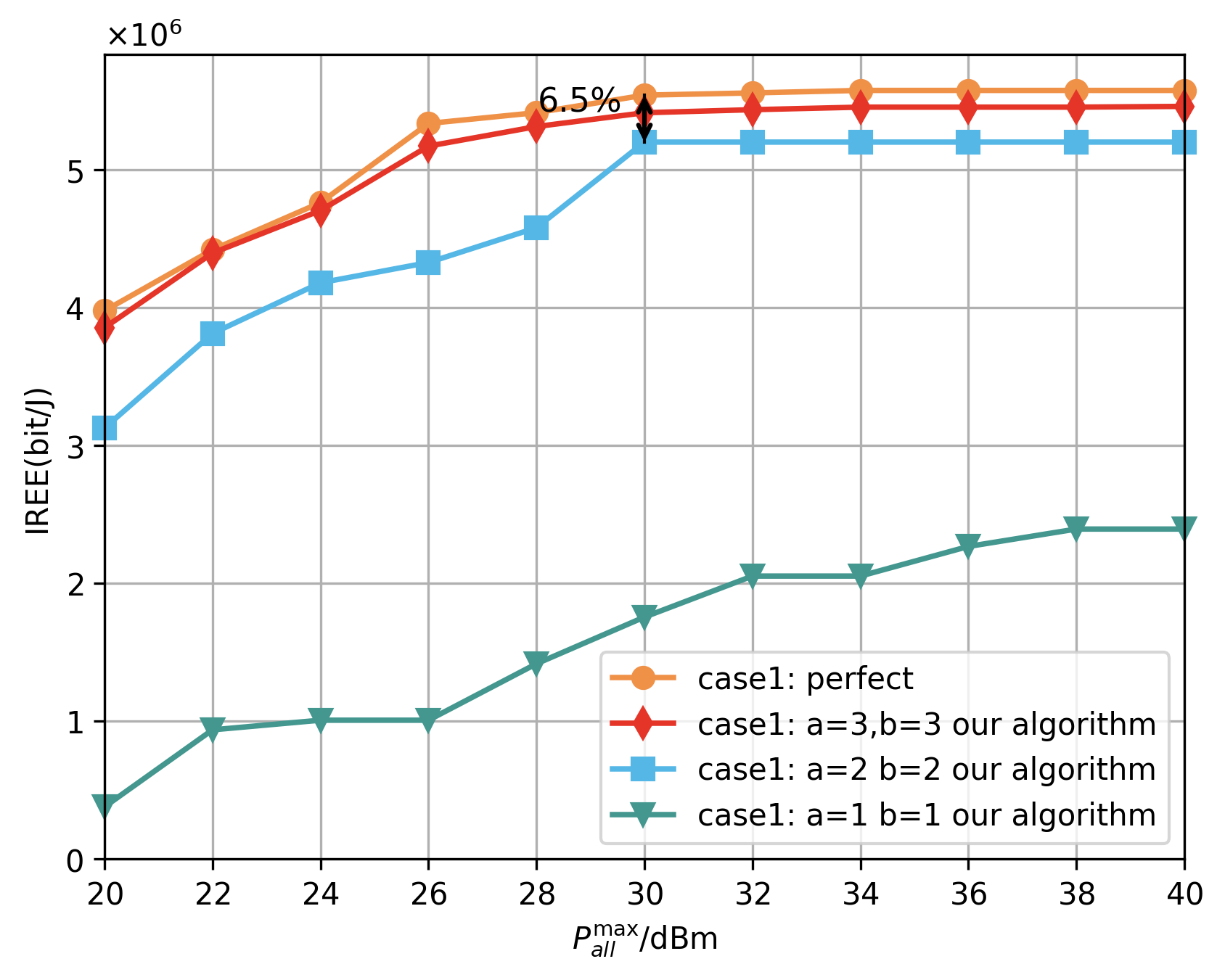} 
    }
    \hfill
    \subfloat[]{
        \includegraphics[width=0.43\textwidth]{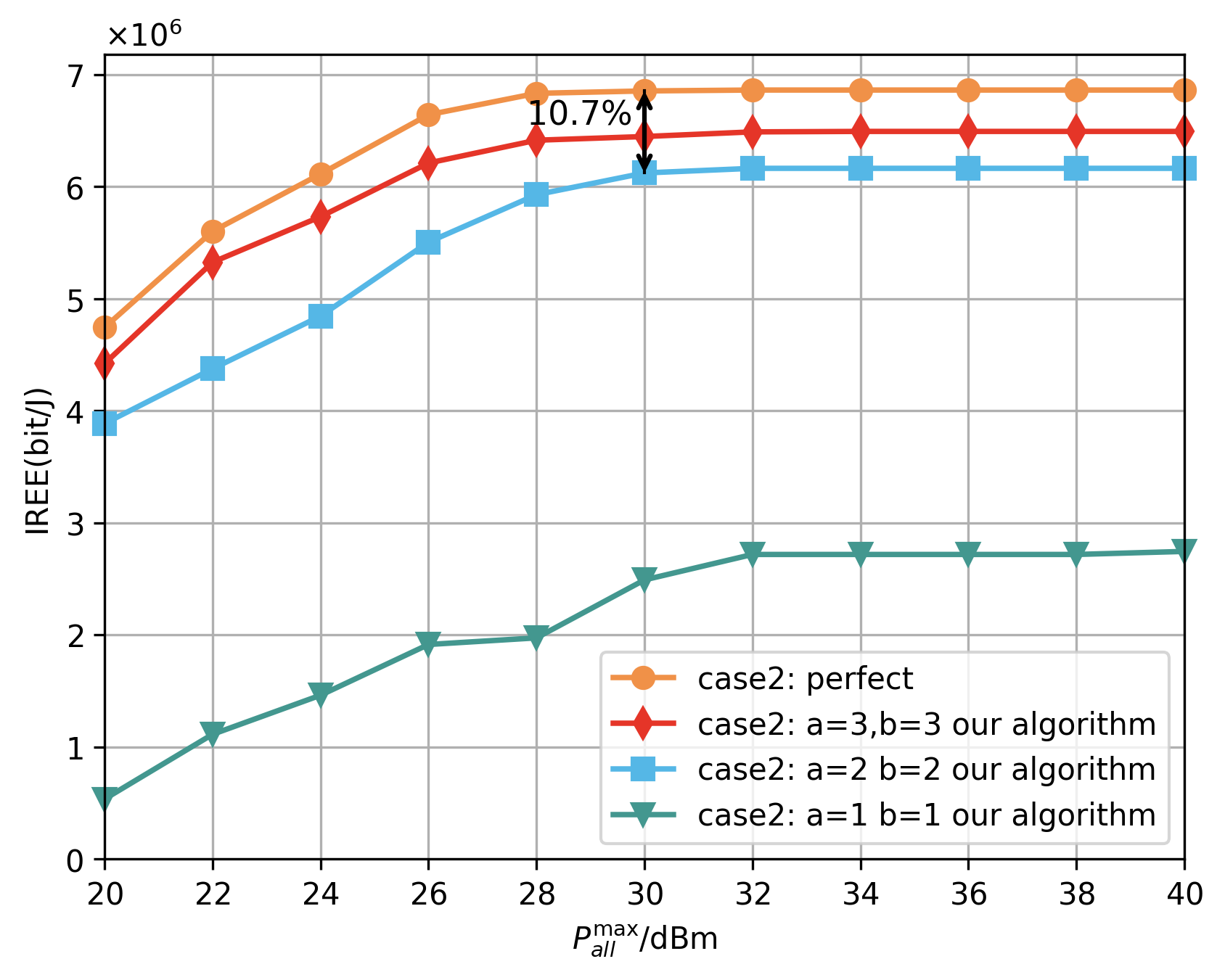} 
    }
    \caption{$\eta_{IREE}$ versus $P_{max}$ in different  quantization levels}
        \label{fig:dif_quan}
    \end{figure*}
The impact of the quantization levels $a$ and $b$ on IREE is shown in Fig. \ref{fig:dif_quan}. For example, at $P^{{\max}}_{all} = 30$ dB, the experiments show that when $a = 2$ and $b = 2$, the performance gap compared to the perfect case is only 6.5\% and 10.7\%. It is noticeable that the disparity is somewhat greater under Case 2 than Case 1. This is primarily due to the higher traffic demand in Case 2, which requires greater energy consumption to compensate for the effects of reflection coefficient discretization. Our method employs multiple approximations to approach the values of $\boldsymbol{\Theta}$ before discretization, ensuring that even if the reflection coefficients at the RIS cannot be continuously adjusted, the optimized IREE value remains very close to the ideal value.
However, in the case of $a=1, b=1$, the performance significantly declines, primarily because the low quantization levels result in poor beamforming at the active RIS. The experiments confirm that, using our algorithm, only modest quantization levels are needed to achieve performance close to the perfect case, which substantially reduces the hardware complexity required for discrete adjustments.

\subsection{The impact of outdated CSI}
\begin{figure*}[!t]
        \centering
        \subfloat[]{
        \includegraphics[width=0.45\textwidth]{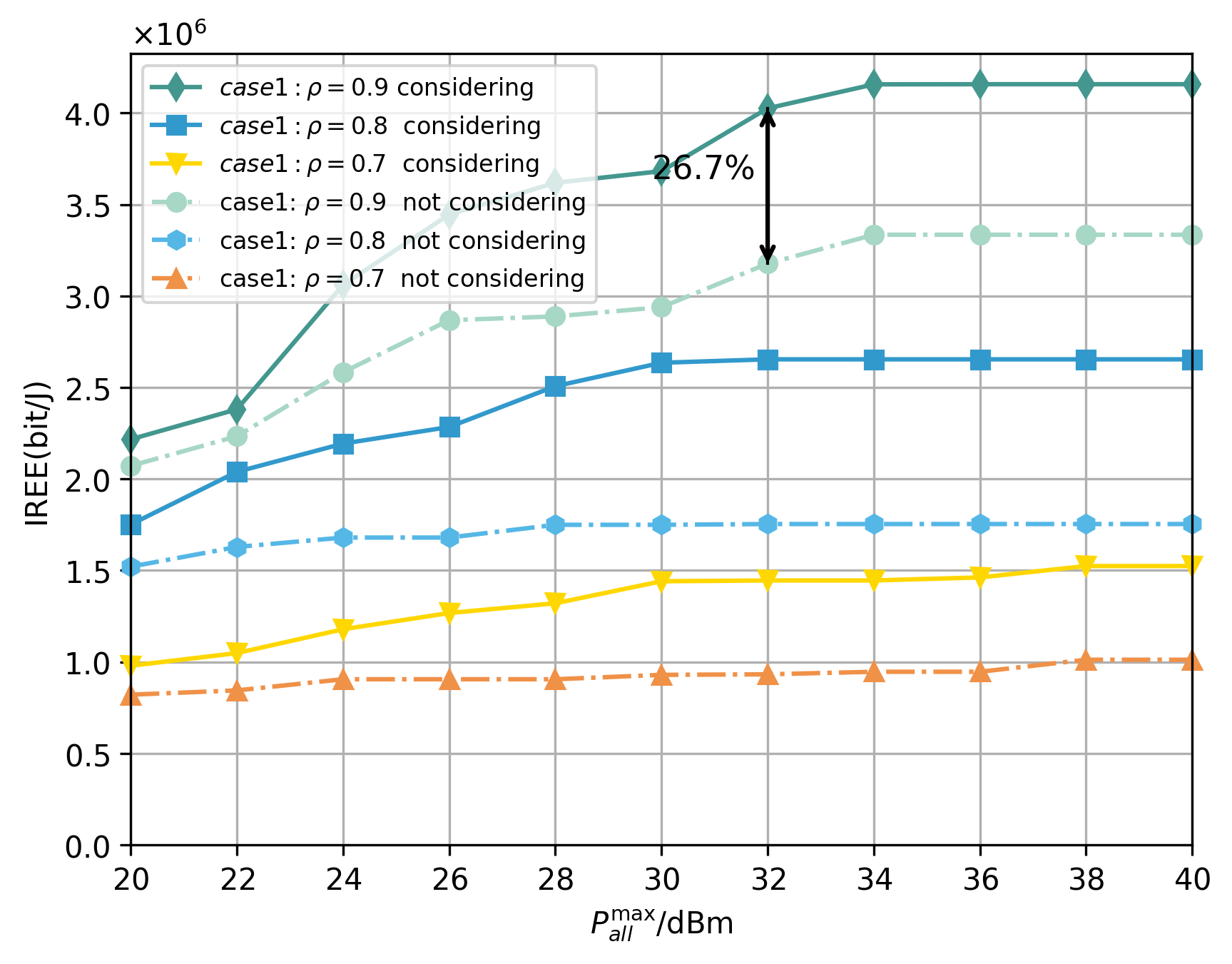} 
    }
    \hfill
   \subfloat[]{
        \includegraphics[width=0.45\textwidth]{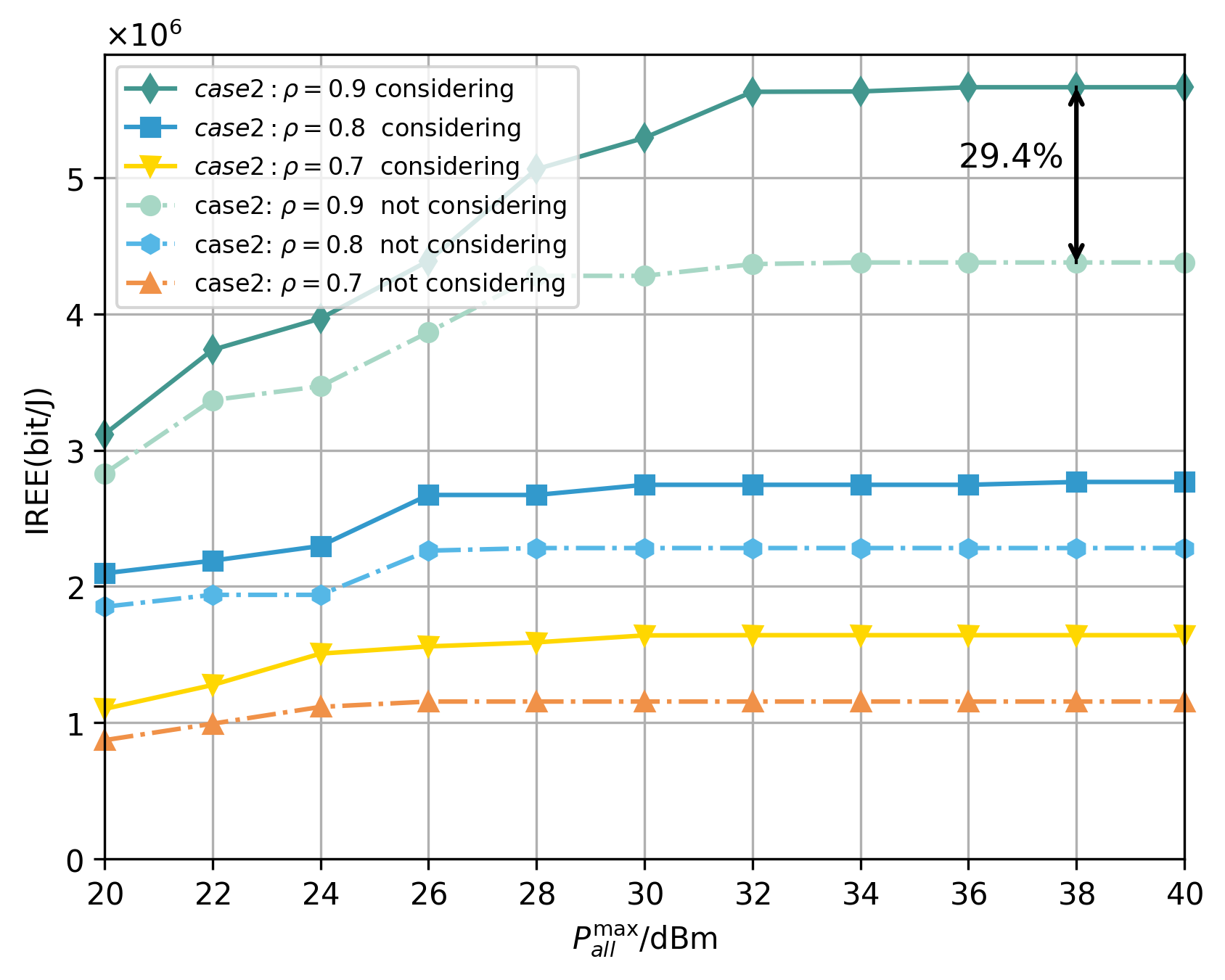} 
    }
    \caption{$\eta_{IREE}$ versus $P_{max}$ in different correlation coefficient $\rho$  }
        \label{fig:dif_csi}
    \end{figure*}
    In Figure \ref{fig:dif_csi}, we illustrate the impact of outdated CSI on IREE. We compare the IREE obtained by optimizing without considering outdated CSI (using $ R_k $) with the IREE obtained by optimizing with outdated CSI (using $ \bar{R}_k $). As $ \rho $ decreases, it becomes evident that the influence of outdated CSI grows more significant, leading to a gradual decline in system performance. This is primarily due to the fact that an increase in outdated CSI introduces additional noise for the users and reduces the power of the useful signal, resulting in a lower SINR, which ultimately degrades overall system performance.

Moreover, it is clear that considering outdated CSI yields a significant performance improvement compared to ignoring it. For instance, at $ \rho = 0.9 $, there are improvements of 26.7\% and 29.4\% in Case 1 and Case 2, respectively. 
This improvement stems from the fact that optimizing without considering outdated CSI fails to effectively mitigate interference from other users and the noise introduced by outdated CSI. This results in a decrease in the actual received SNR, adversely affecting system performance. Furthermore, as $ \rho $ increases, this negative impact becomes even more pronounced. By employing our strategy that incorporates outdated CSI into the optimization process, we can effectively reduce these adverse effects on the system.

 \subsection{IREE improvement compared to EE}
\begin{figure}
        \centering
        \includegraphics[width=0.9\columnwidth]{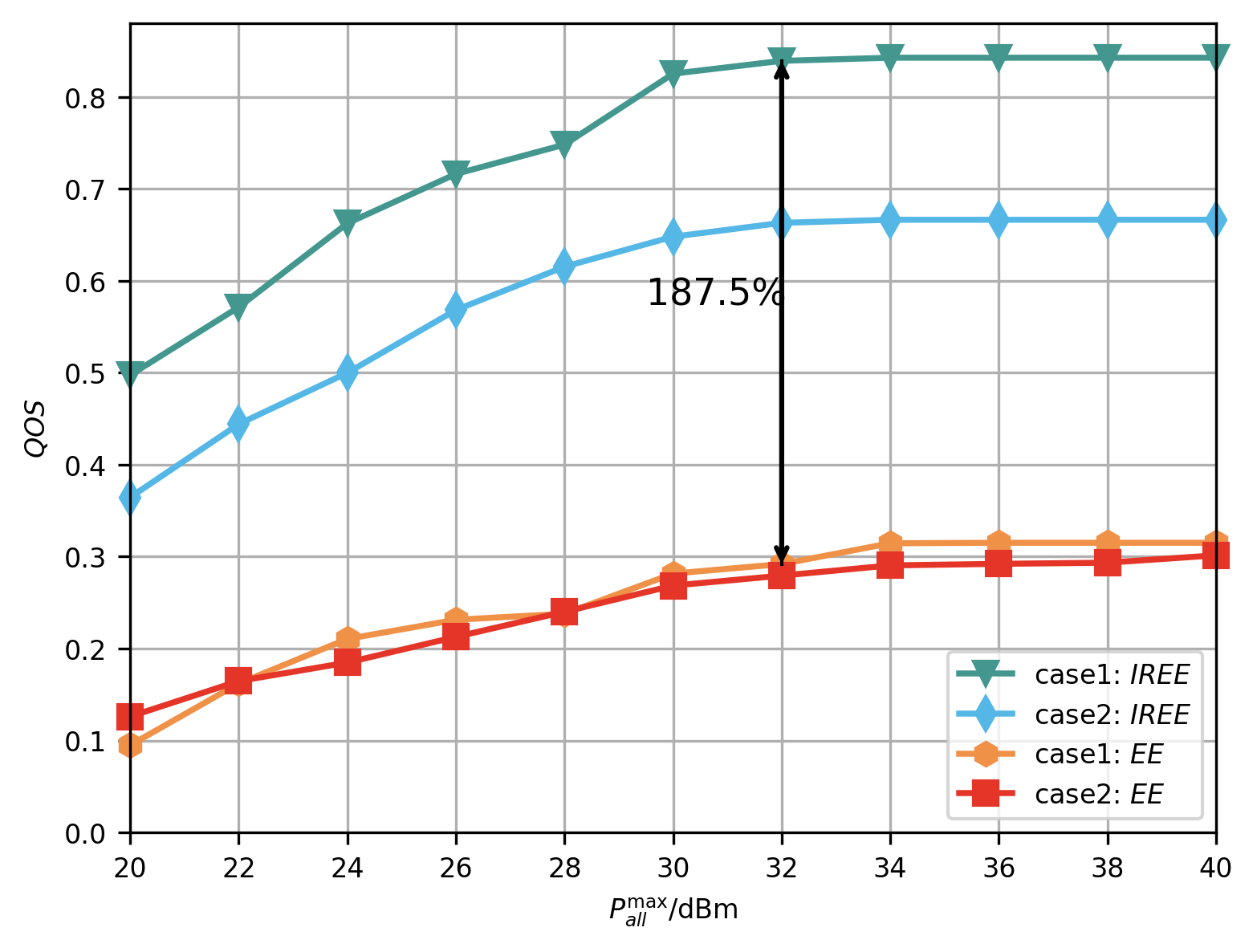}
        \caption{$QOS$ versus $P_{max} $ under the use of EE and IREE}
        \label{fig:IREEVSEE}
\end{figure}
In this section, we use the network utility indicator \cite{yu2024iree} as the quality of service (QoS) metric for our communication system to evaluate the improvement of IREE over the EE metric. This metric allows us to assess the alignment between user demand and actual system capacity. When the actual capacity fully matches user demand, $QoS = 1$; conversely, when there is a complete mismatch, $QoS = 0$.

\begin{align}
    \label{eq:qos}
    QoS &= \frac{\sum_{k=1}^{K} \min \{\bar{R}_k\left(\mathbf{\Theta}, \mathbf{W}\right), D_k\} \xi(\bar{R}, D)}{\sum_{k=1}^{K} D_k} \\
    \xi(\bar{R}, D) &= \frac{1}{2} \sum_{k=1}^{K} [(\bar{R}_k \log_2 ( \frac{2\bar{R}_k}{D_k + \bar{R}_k} ) + D_k \log_2 ( \frac{2D_k}{D_k + \bar{R}_k} )]
\end{align}

where $\xi$ represents the JS divergence \cite{manning1999foundations}. When using the EE metric, we replace IREE in Problem \ref{prob:IREE} with EE for optimization. The specific results are shown in Fig. \ref{fig:IREEVSEE}. The experimental results show that the QoS value increases as $P^{{\max}}_{all}$ increases, eventually reaching an upper limit. Additionally, using IREE significantly improves QoS performance compared to EE. For example, in Case 1, there is an improvement of up to 187.5\%. 

This improvement is mainly because IREE effectively aligns user demand with system capacity. Providing excessive capacity to users with lower demand results in unnecessary energy waste, while users with higher demand require more capacity. In contrast, traditional EE metrics fail to balance user demand and capacity effectively. Furthermore, the QoS performance in Case 1, where user demand is lower, is superior to that in Case 2. This is because, under the same power allocation, it is easier to meet lower user demand, resulting in better QoS.
Moreover, the performance of the EE metric under different traffic conditions is suboptimal. There is almost no improvement in QoS in Case 1 compared to Case 2, indicating that the EE metric fails to effectively match user demands with traffic distribution in both high and low traffic scenarios. In contrast, the IREE metric can better align user demands with traffic, thereby reducing unnecessary energy waste.

\section{Conclusion}
This paper investigated the problem of maximizing  IREE in an active RIS-assisted millimeter-wave communication system based on outdated CSI. Given that perfect CSI between users and the RIS cannot be obtained in practice, we derived the corresponding average achievable rate under outdated CSI conditions. To better meet the demands of green communication, we utilized the network IREE metric based on average achievable rates, which effectively aligns user traffic demands with  network capacity, and formulate the associated IREE optimization framework. An AOSO algorithm was proposed to solve this problem. By employing relaxation and quadratic transform techniques, the original problem is decomposed into two subproblems that can be solved alternately. Additionally, to address the discrete constraints of the RIS reflection coefficients, we introduced a successive approximation method.
Experimental results demonstrate that the proposed algorithm outperforms baseline methods, achieving superior IREE maximization and closely approximating the ideal case of continuous reflection coefficient adjustment. The results also highlighted the importance of optimizing with outdated CSI and show that IREE, compared to traditional EE metrics, better matched user demands with traffic, making it more suitable for the future development of green communication systems.

\begin{appendices} 
\section{Proof of Lemma \ref{le:rk}}
According to Jensen’s inequality , we can make the following deduction \eqref{eq:ergodic_rk} .
\begin{figure*}[hb] 
\centering 
\hrulefill 
\vspace*{8pt} 
\begin{align}
    \label{eq:ergodic_rk}
   \mathbb E\{R_k|&\hat{\mathbf h}_{ru,k}\}=\mathbb E\left[ BW \times \log_2\left(1+\frac{|\rho\hat{\mathbf h}_{ru,k}^H \boldsymbol\Theta \mathbf{H}_{br} \mathbf{w}_k|^2}{\sum^{K}_{k^{'} \neq k} |\mathbf{h}^H_{ru,k} \boldsymbol\Theta \mathbf{H}_{br}\mathbf{w}_{k^{'}}|^2 + ||\mathbf{h}^H_{ru,k}\mathbf\Theta||^2\sigma^2_{R}+\sigma_n^2}\right)\right]\notag\\
    &\geq BW \times \log_2\left(1+\frac{|\rho\hat{\mathbf h}_{ru,k}^H \boldsymbol\Theta \mathbf{H}_{br} \mathbf{w}_k|^2}{\mathbb E\{\sum^{K}_{k^{'} \neq k} |\mathbf{h}^H_{ru,k} \Theta \mathbf{H}_{br}\mathbf{w}_{k^{'}}|^2\} + \mathbb E\{||\mathbf{h}^H_{ru,k}\boldsymbol\Theta||^2\sigma^2_{R}\}+\mathbb E\{|\sqrt{1-\rho^2}\Delta{\mathbf h}_{ru,k}^H \boldsymbol\Theta \mathbf{H}_{br} \mathbf{w}_k|^2\}+\sigma_n^2}\right)
\end{align}
\end{figure*}
Since $\Delta{\mathbf h}_{ru,mk} \sim \mathcal{CN}(0,\sigma^2_{ru,k}\mathbf{I}_{R})$, we can derive that\eqref{eq:enoise}.
\begin{figure*}[hb]
\centering
\hrulefill
\vspace*{8pt}
\begin{align}
\label{eq:enoise}
    \mathbb{E}\{\sum_{k^{'}\neq k}^K |\mathbf{h}_{ru,k}^H \boldsymbol{\Theta} \mathbf{H}_{br} \mathbf{w}_k^{'}|^2\} &= \mathbb{E}\{\sum_{k^{'} \neq k}^K |(\rho\hat{\mathbf h}_{ru,k}^H + \sqrt{1-\rho^2}\Delta\mathbf{h}_{ru,k}^H) \boldsymbol{\Theta} \mathbf{H}_{br} \mathbf{w}_k^{'}|^2\} \notag \\
    &= \sum_{k^{'} \neq k}^K |\rho\hat{\mathbf h}_{ru,k}^H \boldsymbol{\Theta} \mathbf{H}_{br} \mathbf{w}_k^{'}|^2 + \mathbb{E}\{\sum_{k^{'} \neq k}^K |\sqrt{1-\rho^2}\Delta\mathbf{h}_{ru,k}^H \boldsymbol{\Theta} \mathbf{H}_{br} \mathbf{w}_k^{'}|^2\} \notag \\
    &=\sum_{k^{'} \neq k}^K |\rho\hat{\mathbf h}_{ru,k}^H \boldsymbol{\Theta} \mathbf{H}_{br} \mathbf{w}_k^{'}|^2 + |\sqrt{1-\rho^2}\sigma_{ru,k}\mathbf{I}_{R}\boldsymbol{\Theta} \mathbf{H}_{br} \mathbf{w}_k^{'}|^2 \notag \\
    &=\sum_{k^{'} \neq k}^K \mathbf{w}_{k^{'}}^H\mathbf{H}_{br}^H \boldsymbol{\Theta}^H (\rho\hat{\mathbf h}_{ru,k} \hat{\mathbf h}_{ru,k}^H + (1-\rho^2)\sigma_{ru,k}^2 \mathbf{I}_{R}) \boldsymbol{\Theta} \mathbf{H}_{br} \mathbf{w}_k^{'}
\end{align}
\end{figure*}

Let $\mathbf{F}_k = \rho\hat{\mathbf{h}}_{ru,k} \hat{\mathbf{h}}_{ru,k}^H + (1-\rho^2)\sigma_{ru,k}^2 \mathbf{I}_{R}$, a positive definite matrix, which can be decomposed using Cholesky decomposition as $\mathbf{F}_k = \mathbf{G}_k^H \mathbf{G}_k$.  Consequently,
\begin{align}
   \mathbb{E}\{\sum_{k^{'} \neq k}^K |\mathbf{h}_{ru,k}^H \boldsymbol{\Theta} \mathbf{H}_{br} \mathbf{w}_k^{'}|^2\} = \sum_{k^{'} \neq k}^K |\mathbf{G}_k \boldsymbol{\Theta} \mathbf{H}_{br} \mathbf{w}_k^{'}|^2 
\end{align}
Similarly, 
\begin{align}
    \mathbb{E}\left\{|\Delta\mathbf{h}_{ru,k}^H \boldsymbol{\Theta} \mathbf{H}_{br} \mathbf{w}_k|^2\right\} &= ||\sqrt{1-\rho^2}\sigma_{ru,k}\boldsymbol{\Theta} \mathbf{H}_{br} \mathbf{w}_k||^2  \notag\\    
\end{align}
Furthermore,
\begin{align}
    \mathbb{E}\left\{||\mathbf{h}_{ru,k}^H \boldsymbol{\Theta}||^2\right\} &= \mathbb{E}\left\{||\boldsymbol{\Theta}^H \mathbf{h}_{ru,k} \mathbf{h}^H_{ru,k} \boldsymbol{\Theta}||\right\} \notag \\
    &=||\boldsymbol{\Theta}^H\mathbf{F}_{k}\boldsymbol{\Theta}||\notag \\
    &=||\mathbf{G}_{k}\boldsymbol{\Theta}||^2
\end{align}

According to the above proof, we can obtain  the lower bound of  $\mathbb E\{R_k|\hat{\mathbf h}_{ru,k}\}$ which we  define  as the average achievable rate,
\begin{align}
\label{pro:rk}
 \bar{R}_k 
    &= BW \times \log_2\bigg(1+
   \notag \\&\frac{|\hat{\mathbf h}_{ru,k}\boldsymbol\Theta \mathbf{H}_{br}\mathbf{w}_{k}|^2}{\sum_{k^{'}= 1}^K||\mathbf{Q}_{kk^{'}} \boldsymbol\Theta \mathbf{H}_{br}\mathbf{w}_{k^{'}}||^2+||\mathbf{Q}_{k0}  \boldsymbol{\Theta}||^2\sigma_{R}^2 +\sigma_n^2}\bigg) 
    \\
     \mathbf{Q}_{ki} &=
     \begin{cases} 
      \sqrt{\rho\hat{\mathbf h}_{ru,k} \hat{\mathbf h}_{ru,k}^H + (1-\rho^2)\sigma_{ru,k}^2 \mathbf{I}_{R}}&\text{if } k^{'}\neq k. \\
      \sqrt{1-\rho^2}\sigma_{ru,k} & \text{if } k^{'} = k.
     \end{cases}
\end{align}  

\label{appendix:rk}
\section{}
\label{app:1to2}
We let $f(\mathbf W,\boldsymbol \Theta, y)=2y\sqrt{\sum_{k=1}^{K}\min ( \bar{R}_k\left(\mathbf{\Theta}, \mathbf{W}\right), D_k ) }  -y^2 ( P_{BS}(\mathbf{W})+P_{RIS}(\mathbf{\Theta}, \mathbf{W}))$. 
When all variables except $y $ are fixed, $f(\mathbf{W}, \boldsymbol{\Theta}, y) $ becomes a quadratic function, and the optimal value of $y $, denoted as $y^* $, is given by

\begin{equation}
y^* = \frac{\sqrt{\sum_{k=1}^{K} \min \left( \bar{R}_k, D_k \right)}}{P_{BS} + P_{RIS}}.
    \label{eq:y_qd}
\end{equation}

Substituting $y^* $ into $f(\mathbf{W}, \boldsymbol{\Theta}, y) $ yields

\begin{equation}
    f(\mathbf{W}, \boldsymbol{\Theta}, y^*) = \frac{\sum_{k=1}^{K} \min \left\{ \bar{R}_k\left(\mathbf{\Theta}, \mathbf{W}\right), D_k \right\}}{P_{BS}(\mathbf{W}) + P_{RIS}(\mathbf{\Theta}, \mathbf{W})} = \eta_{IREE}.
\end{equation}

Thus, according to \cite{shen2018fractional},we can tackle Problem \ref{prob:IREE2} by iteratively adjusting $y$ while keeping the other variables constant, ultimately facilitating the solution of Problem \ref{prob:IREE}. 

\section{}
\label{app:pro3}
We first prove the equivalence of introducing the slack variable $ \mathbf{u} $. To show that the optimal solution of $ f(\mathbf{W}, \boldsymbol{\Theta}, y) $ is also the optimal solution of $ g(\mathbf{W}, \boldsymbol{\Theta}, y, \mathbf{u}) $, we employ proof by contradiction. Assume the optimal solution of $ \max f(\mathbf{W}, \boldsymbol{\Theta}, y) $ is $ \mathbf{W}^*, \boldsymbol{\Theta}^*, y^* $. Then, there exists $ u_k^* = \min(\bar{R}_k(\mathbf{W}^*, \boldsymbol{\Theta}^*), D_k) ,\forall k \in [1,K] $, such that 
$$
g(\mathbf{W}^*, \boldsymbol{\Theta}^*, y^*, \mathbf{u}^*) = f(\mathbf{W}^*, \boldsymbol{\Theta}^*, y^*).
$$
When $ g(\mathbf{W}, \boldsymbol{\Theta}, y, \mathbf{u}) $ achieves its maximum value, let the optimal solution be $ \mathbf{W}^{'} , \boldsymbol{\Theta}^{'} , y^{'} , \mathbf{u}^{'} $. Then, $ u_k $ must attain its maximum value within its range, i.e., 
$
u_k^{'} = \min(\bar{R}_k(\mathbf{W}^{'} , \boldsymbol{\Theta}^{'}), D_k),
$
at which point 
$$ g(\mathbf{W}^{'}, \boldsymbol{\Theta}^{'}, y^{'}, \mathbf{u}^{'})= f(\mathbf{W}^{'}, \boldsymbol{\Theta}^{'}, y^{'}) $$. 
This leads to 
\begin{align}
g(\mathbf{W}^{'}, \boldsymbol{\Theta}^{'}, y^{'}, \mathbf{u}^{'})&= f(\mathbf{W}^{'}, \boldsymbol{\Theta}^{'}, y^{'}) > g(\mathbf{W}^*, \boldsymbol{\Theta}^*, y^*, \mathbf{u}^*) \notag\\&= f(\mathbf{W}^*, \boldsymbol{\Theta}^*, y^*),\notag
\end{align}
which contradicts our assumption. Therefore, the optimal solution of $ \max f(\mathbf{W}, \boldsymbol{\Theta}, y) $ is indeed the optimal solution of $ \max g(\mathbf{W}, \boldsymbol{\Theta}, y, \mathbf{u}) $. Consequently, we can obtain the optimal solution of $ \max f(\mathbf{W}, \boldsymbol{\Theta}, y) $ by solving $ \max g(\mathbf{W}, \boldsymbol{\Theta}, y, \mathbf{u}) $.
The equivalence of introducing the slack variable $\mathbf{v}$ can be derived similarly. 
   Then when all other variables are fixed except for $\mathbf z$, we have 
\begin{align}
z^*_k = \frac{\textbf{Re}(\mathbf{H}_k \mathbf{w}_k)}{v_k},\forall k \in [1,K].
\end{align}
Substituting $\mathbf z$ into \eqref{eq:c_bs_qd}, we obtain 
\begin{align}
  u_k&\leq\log_2\left(1 + 2 z_k^* \textbf{Re}(\mathbf{H}_k \mathbf{w}_k) - z_k^H v_k z_k^*\right) \notag\\&= \log_2\left(1 + \frac{|\rho \hat{\mathbf{h}}_{ru,k} \boldsymbol{\Theta} \mathbf{H}_{br} \mathbf{w}_{k}||^2}{v_k}\right), \forall k \in [1,K].  
\end{align}
Therefore, we can solve Problem \ref{prob:IREE3} by alternately updating $\mathbf z$ and fixing $\mathbf z$ to optimize the other variables, thereby achieving the objective of solving Problem \ref{prob:IREE2}.

\section{}
\label{app:converge}
The convergence proof for the entire algorithm is presented as follows. Our algorithm fundamentally employs the block coordinate descent (BCD) method \cite{xu2013block} to optimize $ g(\mathbf{W}, \boldsymbol{\Theta}, y, \mathbf{u}, \mathbf{v}, \mathbf{z}) $. After updating each block, the function value of $ g(\mathbf{W}, \boldsymbol{\Theta}, y, \mathbf{u}, \mathbf{v}, \mathbf{z}) $ is monotonically non-decreasing.

Specifically, during the $ i $-th iteration of Algorithm \ref{alg:alter_iree}, since $ g(\mathbf{W}, \boldsymbol{\Theta}, y, \mathbf{u}, \mathbf{v}, \mathbf{z}) $ is a concave function with respect to $ y $ when all other variables are held constant, as well as with respect to $ \mathbf{z} $, we have:
\begin{align}
    &g(\mathbf{W}^{(i-1)}, \boldsymbol{\Theta}^{(i-1)}, y^{(2i-1)}, \mathbf{u}^{(2i-2)}, \mathbf{v}^{(2i-2)}, \mathbf{z}^{(2i-1)}) \notag\\
    &\geq g(\mathbf{W}^{(i-1)}, \boldsymbol{\Theta}^{(i-1)}, y^{(2i-1)}, \mathbf{u}^{(2i-2)}, \mathbf{v}^{(2i-2)}, \mathbf{z}^{(2i-2)}) \notag\\
    &\geq g(\mathbf{W}^{(i-1)}, \boldsymbol{\Theta}^{(i-1)}, y^{(2i-2)}, \mathbf{u}^{(2i-2)}, \mathbf{v}^{(2i-2)}, \mathbf{z}^{(2i-2)}).
\end{align}
Subsequently, since $ g(\mathbf{W}, \boldsymbol{\Theta}, \mathbf{y}, \mathbf{u}, \mathbf{v}, \mathbf{z}) $ is a concave function when $ \mathbf{W}, \mathbf{y}, \mathbf{z} $ are fixed, we obtain the optimal value using CVXPY, leading to:
\begin{align}
\label{eq:prob2}
    &g(\mathbf{W}^{(i-1)}, \boldsymbol{\Theta}^{(i)}, \mathbf{y}^{(2i-1)}, \mathbf{u}^{(2i-1)}, \mathbf{v}^{(2i-1)}, \mathbf{z}^{(2i-1)}) \notag\\
    &\geq g(\mathbf{W}^{(i-1)}, \boldsymbol{\Theta}^{(i-1)}, y^{(2i-1)}, \mathbf{u}^{(2i-2)}, \mathbf{v}^{(2i-2)}, \mathbf{z}^{(2i-1)}).
\end{align}
Similarly, we can derive that:
\begin{align}
\label{eq:prob3}
    &g(\mathbf{W}^{(i)}, \boldsymbol{\Theta}^{(i)}, \mathbf{y}^{(2i)}, \mathbf{u}^{(2i)}, \mathbf{v}^{(2i)}, \mathbf{z}^{(2i)}) \notag\\
    &\geq g(\mathbf{W}^{(i-1)}, \boldsymbol{\Theta}^{(i)}, \mathbf{y}^{(2i)}, \mathbf{u}^{(2i-1)}, \mathbf{v}^{(2i-1)}, \mathbf{z}^{(2i)}).
\end{align}
From these results, we can conclude that:
\begin{align}
g^{(i)}(\mathbf{W}, \boldsymbol{\Theta}, y, \mathbf{u}, \mathbf{v}, \mathbf{z}) \geq g^{(i-1)}(\mathbf{W}, \boldsymbol{\Theta}, y, \mathbf{u}, \mathbf{v}, \mathbf{z}).
\end{align}
Furthermore, due to the power constraints in \eqref{constrain:max_power_bs} and \eqref{constrain:max_power_ris}, $ g(\mathbf{W}, \boldsymbol{\Theta}, y, \mathbf{u}, \mathbf{v}, \mathbf{z}) $ is bounded and non-decreasing, ultimately converging after multiple iterations. Thanks to the equivalence of the quadratic transform, we have:
\begin{align}
\label{eq:prob0}
    \eta_{IREE}^{(i)} = g(\mathbf{W}^{(i)}, \boldsymbol{\Theta}^{(i)}, y^{(2i+1)}, \mathbf{u}^{(2i)}, \mathbf{v}^{(2i)}, \mathbf{z}^{(2i+1)}).
\end{align}
Therefore, the value of IREE  will ultimately converge as $ g(\mathbf{W}, \boldsymbol{\Theta}, y, \mathbf{u}, \mathbf{v}, \mathbf{z}) $ converges. This concludes the proof.  
\end{appendices}

\bibliographystyle{IEEEtran}
\bibliography{IEEEfull,references}
\end{document}